# Bottom-up Synthesis of Metastable 2D Hexagonal Copper(I) Iodide on Monolayer and within Bilayer Graphene

David Kaiser[a], Guobin Jia[b], Janis Köster[c], Sadegh Ghaderzadeh[d], Christian E. Halbig[e], Siegfried Eigler[e], Andrey Turchanin[a], Benjamin Dietzek-Ivanšić[b,f], Arkady V. Krasheninnikov[g], Jonathan Plentz[b], Elena Besley[d], Ute Kaiser[c*]

[a] Institute of Physical Chemistry, Friedrich Schiller University Jena, 07743 Jena, Germany

[b] Leibniz Institute of Photonic Technology (Leibniz-IPHT), 07745 Jena, Germany

[c] Institute for Quantum Optics and Central Facility Materials Science Electron Microscopy, Ulm University, Albert-Einstein-Allee 11, 89081 Ulm, Germany

[d] School of Chemistry, University of Nottingham, University Park, Nottingham NG7 2RD, UK

[e] Institute of Chemistry and Biochemistry, Freie Universität Berlin, Altensteinstraße 23a, 14195 Berlin, Germany

[f] Leibniz Institute of Surface Engineering (IOM), Permoserstr. 15, 04318 Leipzig, Germany

[g] Institute of Ion Beam Physics and Materials Research, Helmholtz-Zentrum Dresden-Rossendorf, 01328 Dresden, Germany.

*Corresponding author: ute.kaiser@uni-ulm.de

**Copper(I) iodide (CuI) is a wide-bandgap semiconductor crystallizing in the 3D *γ*-phase under ambient conditions; its layered van der Waals bulk phase (*β*-CuI) is stable only between 643 and 673 K. The two-dimensional (2D) *h*-CuI form has been obtained via liquid-phase exfoliation of mechanochemically prepared precursors and via encapsulation between graphene sheets, whereas bottom-up growth of 2D *h*-CuI on open surfaces has not yet been demonstrated. Here, we report a vapor-phase synthesis of *h*-CuI directly on low-defect, large-area monolayer and within bilayer reduced oxo-graphene (r-oxo-G) at low temperatures. Using a copper TEM grid as the solid-state precursor for copper, HI-vapor exposure at 40 °C initiates nucleation, while annealing at 180 °C promotes the growth of extended *h*-CuI domains. Aberration-corrected HRTEM resolves the atomic structure, local twist angles, and lattice anisotropy of the CuI/r-oxo-G nanohybrid, while STEM-EDX yields a Cu:I ratio consistent with 1:1. First-principles calculations show that van der Waals adhesion to graphene stabilizes the supported hexagonal layer. Under the presented low-temperature precursor conditions, pathways for nucleation of the *γ*-phase are not available, allowing the hexagonal phase to form selectively at the graphene interface. Ab initio molecular dynamics simulations show that the heterostructure retains its hexagonal lattice order at 600 K, including on an open monolayer graphene support. The lateral extent of the growth is limited mainly by remaining interfacial adsorbates. These results establish a route to metastable 2D *h*-CuI on a chemically inert graphene template, which may be useful for wide-bandgap electronic and optoelectronic devices.**

Two-dimensional (2D) van der Waals (vdW) heterostructures provide a versatile platform for combining materials with different bonding motifs, dimensionalities, and functionalities [1]. In particular, the integration of covalently bonded 2D templates with ionic compounds enables nanohybrid structures that are difficult to realize in bulk form [2–5]. While intrinsically stable 2D materials are now well established [6], increasing attention is being directed toward non-vdW compounds that become metastable in the 2D limit [7,8]. Their stabilization often relies on confinement or interfacial interactions, for example within bilayer graphene or other 2D matrices [4,8–10]. Such approaches have enabled ultrathin silica glass [4], gallium nitride [9], and metal halides [11–13], while graphene encapsulation can additionally protect fragile structures during in situ electron microscopy [14].

Hexagonal copper(I) iodide (*h*-CuI) is the two-dimensional form of copper iodide, identified by Bädeker in 1907 as the first transparent conductor [15]. 2D *h*-CuI combines a wide bandgap, optical transparency, p-type conductivity, and exceptionally low thermal conductivity [16–19], and monolayers have been proposed for transparent electronics, photodetectors, sub-10 nm field-effect transistors, and mechanically compliant wearable devices [17–20]. Under ambient conditions, however, CuI adopts the non-layered zinc-blende *γ*-phase, whereas the layered *β*-phase, the bulk parent of *h*-CuI, is an equilibrium phase only within the narrow temperature range of 643–673 K and reverts to *γ*-CuI on cooling [16,21]. Computational screening predicted a freestanding 2D phase [22], and phonon calculations indicate that it is dynamically stable [17,23]. Thermodynamically, however, the layered stacking lies 3.0–3.4 meV/Å² above *γ*-CuI and is separated from it by a barrier of about 13 meV/Å² for the reconstructive, bond-breaking transformation between the layered and zinc-blende stackings [21]. Thus, *h*-CuI represents a metastable but kinetically trappable 2D phase.

Direct growth of this layered phase on an open surface has remained elusive. Physical vapor deposition from CuI powder yields the cubic *γ*-phase even on van der Waals templates, including $SiO_2$/Si and $WSe_2$/$WS_2$ monolayers [24], while close-distance sublimation on sapphire, Si, GaAs, and GaN produces epitaxial *γ*-CuI containing only a transient 12R polytype embedded in the *γ* matrix [25], a stacking that has been ruled out as the true layered phase [21]. The layered phase has instead been obtained through confinement or top-down processing. Mustonen et al. precipitated *h*-CuI inside bilayer-graphene sandwiches, where the confined gap acts as a nanoreactor [26]; notably, no *h*-CuI formed on the open monolayer regions of the same samples. *β*-CuI nanocrystals have likewise been stabilized within reduced-graphene-oxide membranes [27]. Peng et al. obtained uncovered *h*-CuI flakes by mechanochemical grinding of *γ*-CuI in water followed by liquid-phase exfoliation [23]; these flakes persist under ambient conditions but have no epitaxial relationship to a substrate. Ab initio molecular dynamics simulations further showed that a bare monolayer loses structural integrity at 600 K, whereas graphene-encapsulated 2D *h*-CuI remains stable [28]. Bottom-up growth of the layered phase directly on an open surface has not been reported.

Here, we demonstrate the bottom-up growth of 2D *h*-CuI directly on open monolayer and within bilayer reduced oxo-graphene. The copper TEM grid serves as the solid-state copper source, while HI vapor provides iodine and simultaneously reduces the oxo-graphene template. 80 kV Cc/Cs-aberration-corrected HRTEM resolves the atomic structure, local epitaxial alignment, and uniaxial lattice adaptation of the *h*-CuI/graphene heterostructure, while STEM-EDX establishes a Cu:I ratio consistent with 1:1. DFT calculations quantify the van der Waals adhesion and identify *h*-CuI on an open graphene surface as thermodynamically metastable, whereas AIMD simulations show that the hexagonal lattice is retained up to 600 K. Experimentally, extended growth correlates with removal of interfacial adsorbates, identifying interface cleanliness as a key factor controlling the lateral domain size. These results establish a route to metastable 2D *h*-CuI on a chemically inert graphene template.

# RESULTS AND DISCUSSION

Fig. 1 summarizes the synthesis, a two-step process carried out entirely on a single grid (Methods; Supporting Information, Section S1). The starting material is a film of oxo-functionalized graphene (oxo-G) on a standard copper transmission-electron-microscopy (TEM) grid. In the first step, the film is exposed to hydrogen iodide (HI) vapor at 40 °C, where HI serves a dual purpose: it reduces the oxo-G to reduced oxo-graphene (r-oxo-G) [30], and it attacks the copper of the grid, the solid-state copper source of the synthesis. Since the vapor comes from an aqueous HI solution and the reduction of the oxo-groups releases water, a thin adsorbed film is present at the interface during this step (Section S1). The copper released from the grid crosses the graphene by surface diffusion, on a potential energy landscape flat enough for copper adatoms to migrate anomalously at room temperature [32], and meets the iodine adsorbed during the HI treatment, where small hexagonal CuI (*h*-CuI) crystallites nucleate. The copper originates solely from the grid: a control synthesis on a gold (Au) TEM grid, under otherwise identical conditions, yields no *h*-CuI (Supporting Information, Section S1). In the second step, annealing at 180 °C grows the crystallites into extended, van der Waals-epitaxial domains. Every step takes place on the original graphene substrate, so the route needs no transfer step and no sacrificial polymer; the same two-step route also forms *h*-CuI within bilayer graphene, by intercalation. The growth mechanism and the selection of the hexagonal phase are discussed in detail below.

**Template characterization and substrate morphology**

The structural integrity and morphology of the oxo-G template were first examined prior to HI treatment and thermal annealing. Overview TEM imaging [Fig. 2(a)] confirms the formation of continuous, large-area oxo-G membranes spanning the copper TEM grid support. 80 kV $C_c/C_s$-corrected high-resolution TEM (HRTEM) reveals a heterogeneous substrate morphology consisting of monolayer, bilayer, and trilayer regions. A representative boundary between monolayer and trilayer areas is shown in Fig. 2(b), with the corresponding layer numbers confirmed by the fast Fourier transform (FFT) patterns shown in the insets.

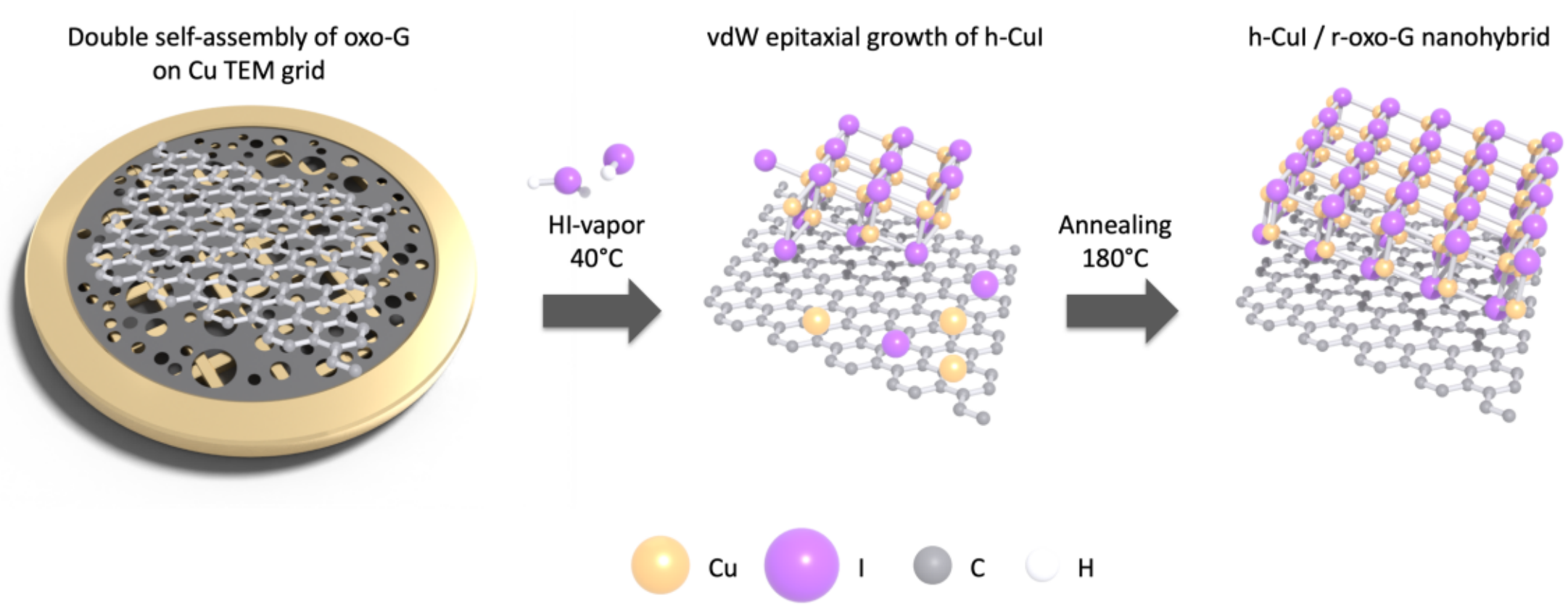

**Figure 1. *h*-CuI synthesis on graphene supported by a copper TEM grid.** In the first step, HI treatment at 40 °C forms, through van der Waals stabilization, monolayer *h*-CuI on graphene; during this step the oxo-graphene template (oxo-G) is reduced to r-oxo-G simultaneously. Annealing at 180 °C then promotes the epitaxial growth of extended *h*-CuI crystals on graphene. The atom types are indicated at the bottom.

High-resolution lattice images of monolayer and bilayer regions [Figs. 2(c) and 2(e)] demonstrate pronounced long-range crystalline order within the oxo-G sheets, despite minor amorphous surface residues that are typical of liquid-phase exfoliation and transfer processes. In bilayer regions, rotational misalignment between the two constituent sheets gives rise to distinct moiré superlattices [Fig. 2(e)]. The FFT analysis in Fig. 2(f) shows two discrete sets of sharp hexagonal reflections, labeled G and G′, corresponding to a relative twist angle of 16.7 ± 1.0°. These well-defined crystallographic signatures demonstrate that the oxo-G retains the lattice periodicity required to act as a robust template for van der Waals (vdW) epitaxy. The coexistence of monolayer and bilayer regions from the same exfoliation batch on the same grid further enables a direct comparison of *h*-CuI growth on SLG and BLG.

### Vapor-phase growth on monolayer graphene: HRTEM analysis

We first examined *h*-CuI formation on samples that received HI treatment at 40 °C without subsequent annealing. HRTEM imaging revealed the nucleation of discrete *h*-CuI crystallites of about 5 nm on monolayer and bilayer r-oxo-G. Fig. 3(a) shows a crystallite epitaxially grown on monolayer r-oxo-G. FFT analysis of the adjacent substrate confirms the monolayer nature of the graphene template [inset, Fig. 3(a)], while the heterostructure FFT [Fig. 3(b)] shows the white graphene reflections G together with a distinct set of red reflections from *h*-CuI. Because the small lattice mismatch can bring an *h*-CuI reflection close to a graphene spot, the layer number is independently confirmed at an adjacent graphene-only region [inset, Fig. 3(a)]. This region shows a single set of first-order graphene reflections with no moiré pattern, which identifies the substrate as monolayer. The *h*-CuI lattice vectors $b_1$, $b_2$, $b_3$ are determined from the HRTEM image by fitting two-dimensional Gaussians to the atomic columns [31], while the graphene primitive vectors $a_1$, $a_2$, $a_3$ are obtained from the corresponding FFT and serve as the internal length reference; all measured lattice vectors and twist angles are tabulated in Table S1. In every annealed crystallite the largest of the three lattice vectors reaches the commensurate graphene superlattice value of $a\sqrt{3} \approx 0.426$ nm, with 0.426–0.427 nm, and the smallest stays at the freestanding *h*-CuI value of 0.419 nm, with 0.418–0.421 nm — each within the measurement uncertainty of ± 0.002 nm (Table S1). This pattern holds on monolayer and within bilayer graphene alike; the adaptation of the individual crystallites is therefore characterized below from their anisotropy. The epitaxial twist angle is defined as $\theta = |30° - \Delta|$, where Δ is the angle between the *h*-CuI lattice vector $\mathbf{b}_3$ and the nearest graphene primitive vector a. Physically, the 30° offset reflects the energetically preferred registry in which the iodine atoms occupy the centers of the graphene hexagons [26]; this hexagon-center sublattice is rotated by 30° relative to the graphene atomic lattice, so θ = 0 corresponds to perfect epitaxial lock-in (Section S2c). For the nucleus shown in Fig. 3, this yields θ = 11.9° ± 1.0°. Thermal annealing at 180 °C significantly improves crystallinity and promotes lattice relaxation. Post-annealing HRTEM imaging [Fig. 3(d)] reveals increased domain sizes and sharper diffraction spots [Fig. 3(e)], which again show the *h*-CuI reflections together with those of graphene. The monolayer assignment is confirmed again by a separate windowed FFT of an adjacent graphene-only region [inset, Fig. 3(d)], which shows a single set of first-order graphene reflections without a moiré pattern, ruling out a bilayer. The measured lattice vectors $b_1 = 0.421 \pm 0.002$ nm, $b_2 = 0.425 \pm 0.002$ nm and $b_3 = 0.427 \pm 0.002$ nm span the range between the freestanding and the commensurate value.

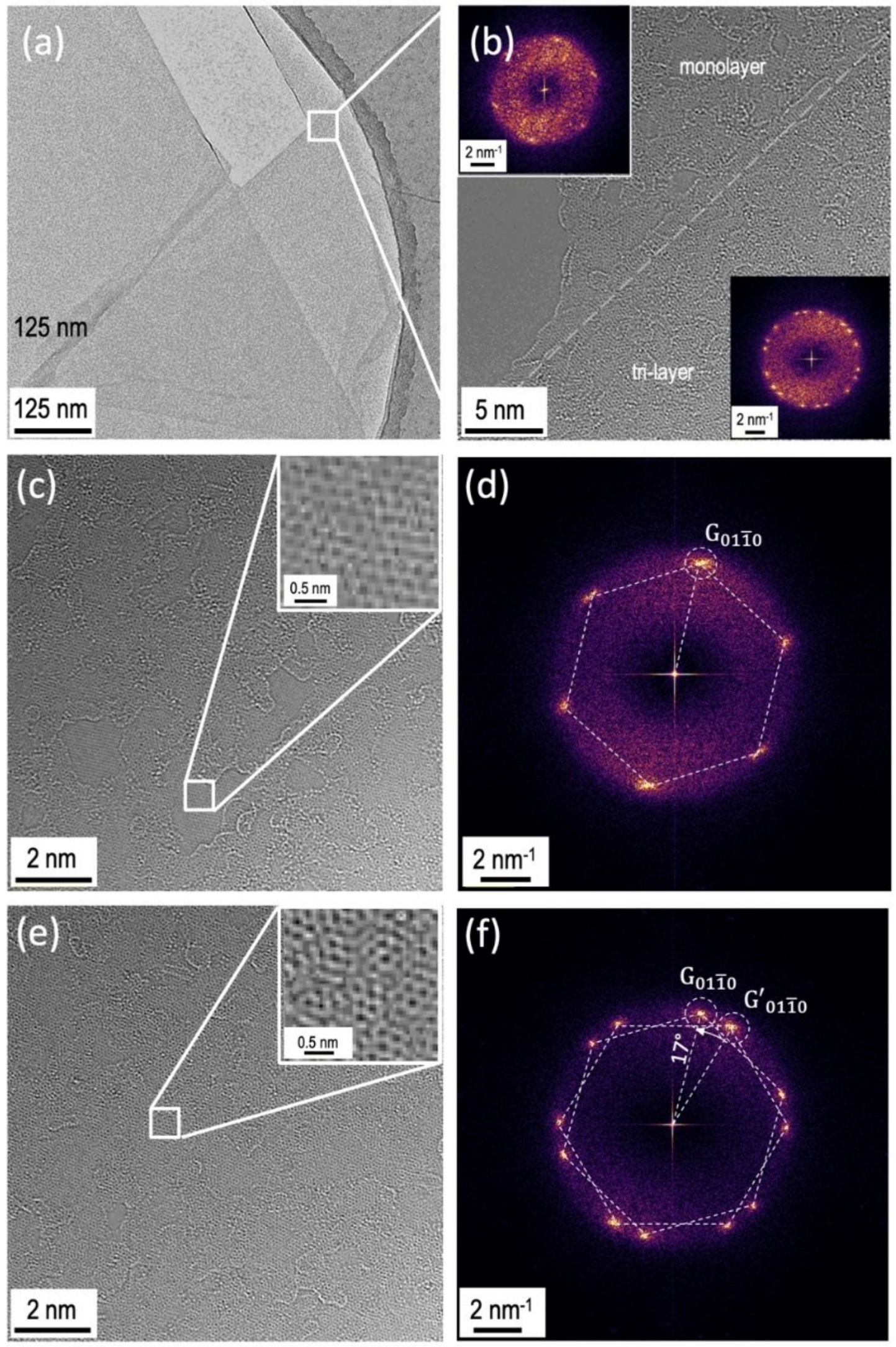

**Figure 2. Structural and crystallographic characterization of oxo-G templates.** (a) Low-magnification $C_c/C_s$-corrected HRTEM overview of oxo-G flakes suspended on a TEM grid, showing characteristic folds and local thickness variations. (b) $C_c/C_s$-corrected HRTEM image of the interface between a monolayer domain (top) and a trilayer domain (bottom), indicated by the white line; the insets show the corresponding FFT patterns. (c) Representative $C_c/C_s$-corrected HRTEM image of a monolayer oxo-G region; the inset shows an atomic-resolution view of the hexagonal lattice. (d) FFT pattern obtained from the region shown in (c), displaying the reflections of a single graphene lattice, labeled G and marked by the dashed circle. (e) $C_c/C_s$-corrected HRTEM image of a rotationally misaligned bilayer region exhibiting a moiré superlattice, magnified in the inset. (f) Corresponding FFT pattern of the bilayer region, showing two sets of hexagonal reflections, labeled G and G′, with a relative twist angle of 16.7 ± 1.0°. All data were acquired at an accelerating voltage of 80 kV.

To characterize how the *h*-CuI lattice adapts to the graphene template, we compare the measured lattice vectors with the commensurate graphene superlattice period ($a\sqrt{3} \approx 0.426$ nm). The adaptation is uniaxial: along one direction an *h*-CuI vector matches the substrate period, while the perpendicular vector stays close to the intrinsic *h*-CuI value and retains a residual mismatch of about 1.4% (from the initial 1.67% lattice mismatch). Across the measured crystals this leaves a small directional anisotropy — the ratio of the longest to the shortest lattice vector — of 1.014–1.019 (Table S1). In our open monolayer the perpendicular direction is not contracted below the intrinsic *h*-CuI value, i.e. the uniaxial adaptation proceeds without a perpendicular (Poisson) compression, whereas Mustonen et al. [26] report a comparable anisotropy for graphene-encapsulated 2D *h*-CuI. The energetics that select this uniaxial adaptation — the registry (corrugation) energy gained on commensuration against the elastic strain cost, and the lower bound it places on the interfacial corrugation — are analyzed in detail in the Growth mechanism and phase selection section.

**Intercalation and growth within bilayer graphene**

The presence of both monolayer and bilayer graphene flakes in the oxo-G substrate enables a direct comparative study of *h*-CuI growth under identical experimental conditions. Following HI treatment at 40 °C, *h*-CuI nucleation is also observed within bilayer regions. HRTEM imaging [Fig. 4(a)] reveals intercalated *h*-CuI crystallites between two graphene sheets. The bilayer nature is confirmed by the FFT of the adjacent region [inset, Fig. 4(a)], showing the two hexagonal graphene lattices G and G′ with a relative twist angle of 26.0° ± 1.0°. The heterostructure FFT [Fig. 4(b)] displays triple-lattice symmetry from G, G′, and the red *h*-CuI reflections. For this specific crystallite, *h*-CuI shows epitaxial deviation angles of θ = 14.4° ± 1.0° to G and θ′ = 19.6° ± 1.0° to G′. Atomic position mapping yields lattice vectors of $b_1$ = 0.416 ± 0.002 nm, $b_2$ = 0.416 ± 0.002 nm, and $b_3$ = 0.419 ± 0.002 nm, corresponding to $A_b$ = 1.007 ± 0.004. These values correspond to an essentially isotropic crystallite. We attribute the near-absence of in-plane anisotropy to the early stage of growth: the nucleus is small and has a high edge-to-area ratio, so its lattice has not yet developed the directional adaptation that characterizes larger, fully grown domains. The extracted constants therefore remain close to the intrinsic 2D *h*-CuI value of 0.419 ± 0.002 nm. Thermal annealing at 180 °C promotes the further growth of these intercalated domains [Fig. 4(d)]. In the observed samples, *h*-CuI often achieved near-perfect alignment with one graphene layer, here G′, for which θ′ ≈ 0° and the diffraction spots are rotated by 30°, while it maintained a twist angle of θ = 21.7° ± 1.0° with the other layer, G [Fig. 4(e)]. The obtained lattice parameters are b1 = 0.419 ± 0.002 nm, b2 = 0.422 ± 0.002 nm, and b3 = 0.426 ± 0.002 nm. The anisotropy of 1.7% indicates the uniaxial strain profile, reflecting the lattice adaptation to the epitaxial template in only one direction, while *h*-CuI remains relaxed in the perpendicular direction, where b1 = 0.419 ± 0.002 nm is close to the intrinsic value. The simultaneous formation of *h*-CuI in monolayer and bilayer regions of the same substrate, under identical conditions, enables a direct comparison of the two configurations. In both, *h*-CuI grows epitaxially with a hexagonal lattice, which shows that intercalation into twisted bilayer graphene proceeds analogously to growth on the open monolayer surface. An additional crystal, which is located within the same bilayer flake and which also shares a grain boundary with the crystal of Fig. 4(d,e), is shown in Fig. S3. Yet it is locked to G instead of G′. A crystallite on an outer surface of bilayer graphene would be templated by the outermost sheet in both cases, so registry with different graphene layers would be impossible. Both crystals are shown together in one image in Fig. S2, identifying the *h*-CuI as encapsulated within two graphene sheets (Section S3).

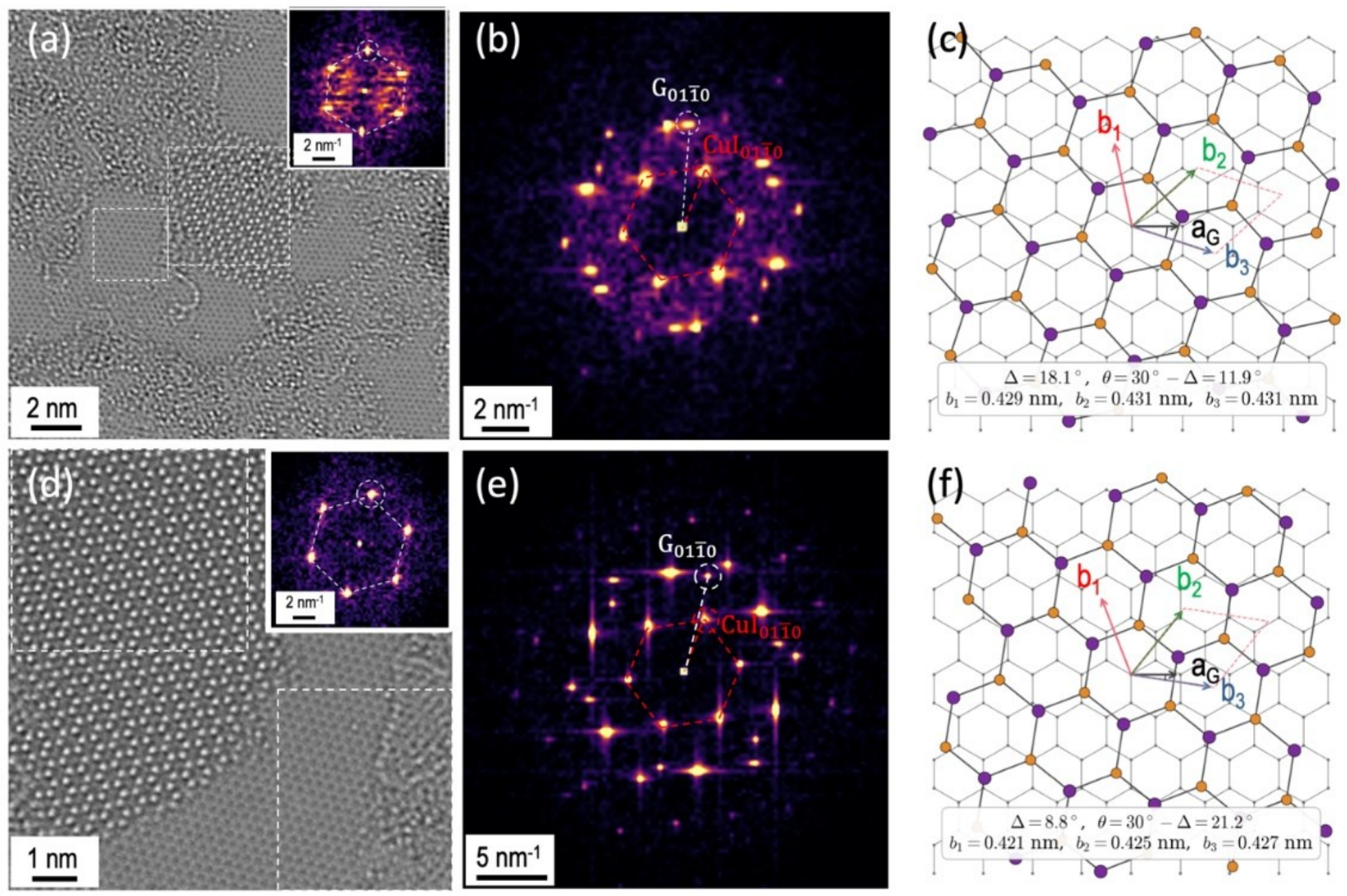

**Figure 3. Vapor-phase synthesis and vdW-epitaxial growth of *h*-CuI on monolayer graphene.** (a) $C_c/C_s$-corrected HRTEM image (80 kV) of an *h*-CuI crystallite (~5 nm) nucleated on monolayer graphene following HI treatment at 40 °C. The inset FFT of the adjacent substrate confirms the monolayer nature of the graphene template (G). (b) FFT of the heterostructure region displaying commensurate reflections from graphene (G, white) and *h*-CuI (red circles). (c) Schematic of the epitaxial registry at 40 °C: *h*-CuI on SLG with iodine atoms occupying graphene hollow sites at a twist angle of θ = 11.9° ± 1.0° (Sample 1). Views are reconstructed from experimental lattice vectors ($b_1$, $b_2$, $b_3$) and measured twist angles. (d) HRTEM image of an expanded *h*-CuI domain (up to about 30 nm) on monolayer graphene after thermal annealing at 180 °C. (e) Corresponding FFT showing enhanced intensity and sharpness of *h*-CuI reflections, indicating improved crystalline quality and lattice relaxation post-annealing (θ = 21.2° ± 1.0°, Sample 2). (f) Structural schematic of the annealed heterostructure, illustrating the evolution of the moiré superlattice.

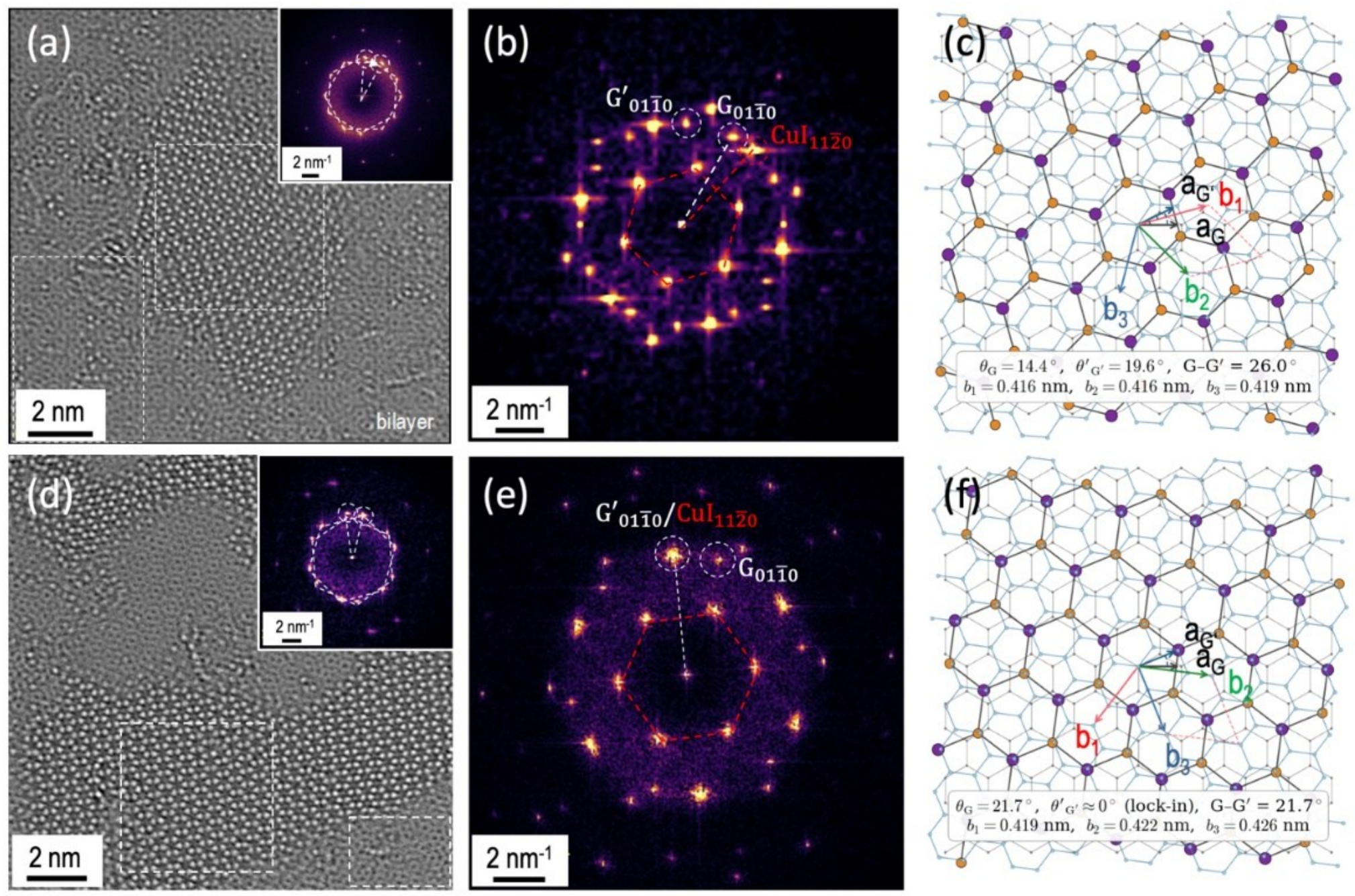

**Figure 4. Intercalation and growth of *h*-CuI within twisted bilayer graphene.** (a) $C_c/C_s$-corrected HRTEM image (80 kV) of an *h*-CuI crystallite intercalated between two graphene layers after HI treatment at 40 °C. The inset FFT of the adjacent region reveals two hexagonal graphene lattices (*G*, *G′*) with a relative twist angle of 26.0° ± 1.0°, confirming the bilayer template. (b) FFT of the triple-layer heterostructure displaying reflections from both graphene layers (*G*, *G′*, white) and the intercalated *h*-CuI (red circles). (c) Structural schematic of the 40 °C configuration (Sample 3): *h*-CuI is sandwiched between *G* and *G′* with epitaxial twist angles θ = 14.4° and θ′ = 19.6° (defined as the deviation from 30° epitaxy, θ = |30° − Δ|, with Δ taken against the corresponding graphene layer; the raw lattice-vector angles Δ = 15.6° and Δ′ = 10.4° are listed in the SI), the two raw angles adding to Δ + Δ′ = 26.0° and thus reproducing the G–G′ twist of 26.0° ± 1.0° read directly from the FFT in panel (a). (d) HRTEM micrograph showing an expanded *h*-CuI domain within a bilayer region after thermal annealing at 180 °C. (e) Corresponding FFT showing *h*-CuI reflections (red) in near-epitaxial alignment with one graphene layer (θ′ ≈ 0°, locked to G′), while maintaining a twist angle of θ = 21.7° with the other layer (G). The overlapping reflection is the second-order *h*-CuI reflection, which nearly coincides with the first-order graphene spot by lattice commensurability. (f) Schematic reconstruction of the annealed *h*-CuI/BLG heterostructure based on experimental lattice vectors and orientation mapping.

**Stoichiometric and elemental analysis**

To verify the chemical identity and spatial distribution of the synthesized crystals, we performed STEM-EDX analysis following the 180 °C annealing step. High-angle annular dark-field (HAADF) imaging [Fig. 5(a)] — complemented by elemental mapping, see Supporting Information [Fig. S1] — reveals extended areas of *h*-CuI over several hundred nanometers on the graphene membrane. The magnified view [Fig. 5(b)] resolves these *h*-CuI islands at higher contrast, on both monolayer and few-layer graphene regions, further demonstrating that our vapor-phase route facilitates extensive growth on both open surfaces and within encapsulated environments.

The composite EDX elemental map [Fig. 5(c)] shows copper (Cu) and iodine (I) co-localized in the regions where crystals have formed, while the carbon (C) signal remains uniform across the graphene template. In the crystal-free region [region 1 in Fig. 5(c)] the Cu and I fractions fall to 0.38 and 0.05 at%, that is, 18 and 102 times below the crystal-covered region. Individual elemental maps are provided in Fig. S1 (Supporting Information). Quantitative EDX spectroscopy of the crystal-covered region [region 2 in Fig. 5(c); Fig. 5(d)] yields Cu and I atomic fractions of 7.00 ± 1.07% and 5.10 ± 0.67%. The resulting Cu:I ratio of 1.4 ± 0.3 is consistent with the 1:1 stoichiometry of copper(I) iodide within the accuracy of standardless EDX quantification. Part of the copper signal originates outside the crystals: the crystal-free region records 0.38 at% Cu. The bright, high-contrast particles in the HAADF images stem from residual gold contamination, located mainly on the holey-carbon support (Supporting Information, Section S1).

Thermal annealing significantly reduces the oxygen signal in the EDX maps (see Fig. S1e), indicating the effective removal of oxygen-containing functional groups and adsorbed water from the initial oxo-G template during the HI treatment and subsequent annealing. Sulfur (from the SDS surfactant) and sodium were not detected within the signal-to-noise limits of the current EDX acquisition. This suggests an efficient removal of SDS surfactant used during the DSA process to form the continuous oxo-G film. The absence of these elements confirms that the synthesis provides a clean graphene template for the formation of large-area *h*-CuI/graphene heterostructures. A weak Ca K line at 3.7 keV is also visible in the spectrum. It was not included in the quantification and is attributed to residues from the sample preparation.

**Growth mechanism and phase selection**

Next, we discuss the possible growth mechanisms and the formation energies of the uniaxially strained *h*-CuI phase on SLG. An estimate of the formation energies indicates that *h*-CuI is metastable. The energetic penalty of the freestanding 2D phase relative to bulk γ-CuI [21] is estimated to be about 20.5 meV/Å² (see SI Section S4c). We performed DFT calculations to obtain $E_{vdW}$ for the untwisted 3 × 3 *h*-CuI on 5 × 5 graphene supercell, see the supporting information Sections S4a–S4b. The van der Waals adhesion $E_{vdW}$ of the heterostructure is 13.58 meV/Å$^2$ (0.218 J/m$^2$) per graphene interface (SI Section S4c). This value is slightly lower than the vdW binding energy of 17.4 meV/Å$^2$ for the CuI–CuI interlayer. Thus, the *h*-CuI monolayer on SLG is energetically about 7.0 meV/Å$^2$ less favorable than γ-CuI. A graphene-encapsulated *h*-CuI monolayer would be more favorable than γ-CuI by 6.6 meV/Å$^2$. The *h*-CuI monolayer on a graphene substrate is therefore metastable, while a second graphene interface would make it thermodynamically favored.

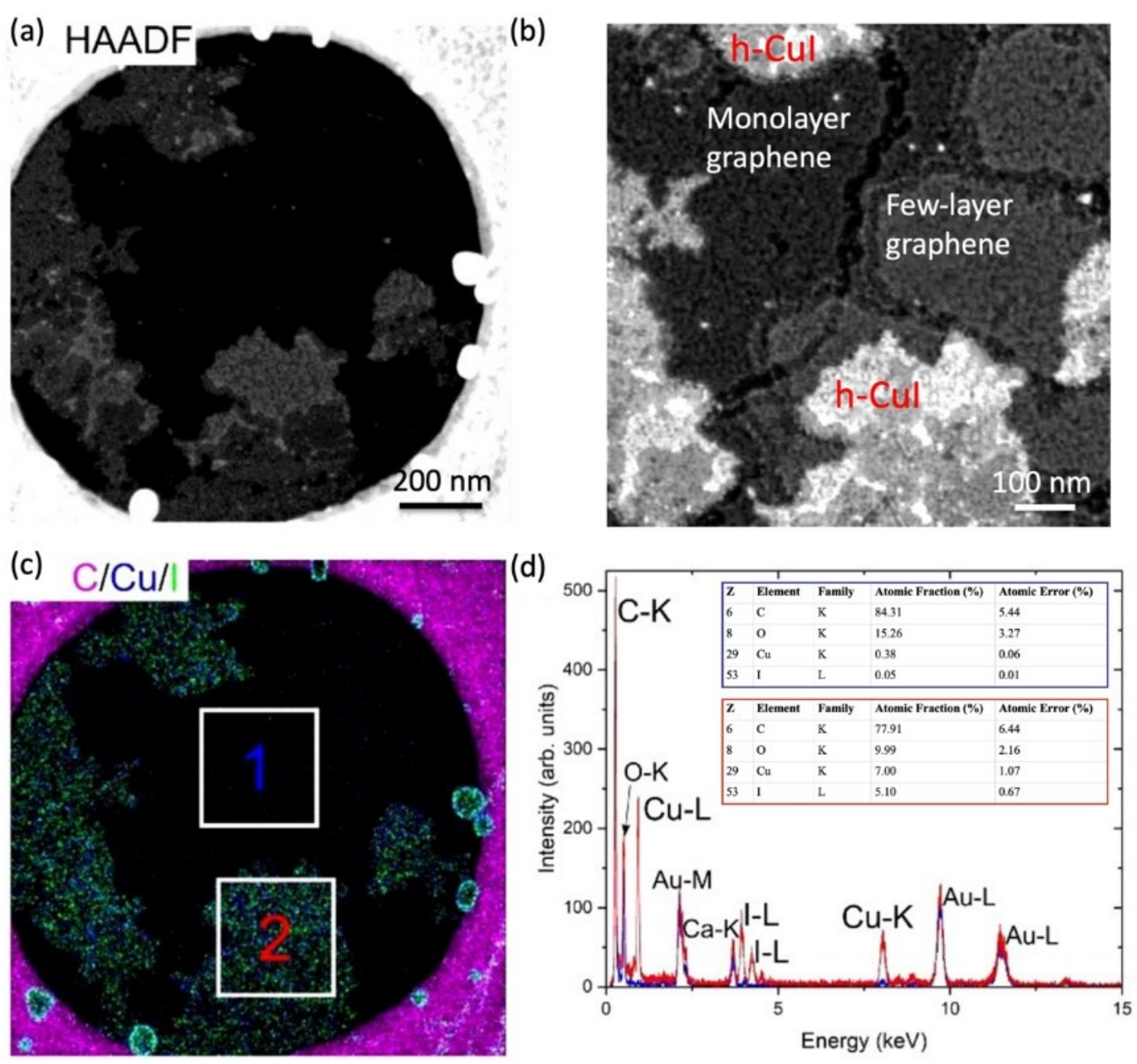

| Z | Element | Family | Atomic Fraction (%) | Atomic Error (%) |
|---|---|---|---|---|
| 6 | C | K | 84.31 | 5.44 |
| 8 | O | K | 15.26 | 3.27 |
| 29 | Cu | K | 0.38 | 0.06 |
| 53 | I | L | 0.05 | 0.01 |

| Z | Element | Family | Atomic Fraction (%) | Atomic Error (%) |
|---|---|---|---|---|
| 6 | C | K | 77.91 | 6.44 |
| 8 | O | K | 9.99 | 2.16 |
| 29 | Cu | K | 7.00 | 1.07 |
| 53 | I | L | 5.10 | 0.67 |

**Figure 5. Elemental and stoichiometric characterization via STEM-EDX.** (a) HAADF-STEM overview of the graphene flakes spanning the holey-carbon support, with extended areas of *h*-CuI; the bright particles are residual gold contamination (Supporting Information, Section S1). (b) Magnified view of the area in (a) at higher contrast, showing *h*-CuI domains (labeled) on both monolayer and few-layer graphene regions. (c) Composite EDX map of carbon (magenta), copper (blue), and iodine (green); Cu and I co-localize in the crystal-covered regions, whereas in the crystal-free region 1 both fall by one to two orders of magnitude. (d) Representative EDX spectrum acquired from the *h*-CuI domain labeled “2” in (c), giving a Cu:I atomic ratio consistent with 1:1 within the quantification accuracy. Inset tables provide a comparative stoichiometric analysis of the crystal and the bare graphene substrate. The essential role of the Cu grid as the sole solid-state precursor was independently verified by a control experiment using an Au TEM grid, which yielded no *h*-CuI under otherwise identical conditions (see SI Section S1, Sample Preparation).

Its formation, however, is governed not only by thermodynamics but also by the available formation routes. First, the formation of γ-CuI would require a three-dimensional nucleus with dangling bonds. In classical nucleation theory the barrier for such a three-dimensional nucleus scales as ΔG* ∝ γsurf3/Δμ2 (γsurf the surface free energy, Δμ the chemical-potential driving force). Under our conditions, the precursor elements arrive separately at a graphene interface. Iodine resides on the graphene surface after the HI treatment, and copper migrates in from the grid bars over a potential energy landscape flat enough to permit anomalous adatom diffusion at room temperature [32]. Thus, CuI forms in situ under dilute conditions, and the local supersaturation required to nucleate a three-dimensional *γ*-CuI particle is unlikely to be reached. Furthermore, routes that yield *γ*-CuI nanosheets and epitaxial *γ*-CuI films sublime CuI powder from sources held at 360–450 °C [24,25]; in our process no component exceeds 180 °C, far below appreciable CuI sublimation, so the molecular CuI flux that enables *γ*-phase growth is absent. Second, the *h*-CuI surface is self-saturating: its basal planes carry no dangling bonds, so the iodine-terminated basal *h*-CuI plane cannot template *γ*-CuI. Moreover, the precursor species Cu and I are supplied on graphene and not on top of *h*-CuI, so a second layer cannot nucleate, which makes multilayer *h*-CuI (*β*-CuI) equally unlikely. Under these conditions where HI functions as both the reducing agent and the iodine source, the hexagonal phase does not have to be thermodynamically more favorable than *γ*-CuI. The conditions under which *γ*-CuI nucleates do not arise in the first place.

The observation of uniaxial strain in the extended *h*-CuI crystals (Samples 2, 4, and 5, and Ref. [26]) provides a constraint on the total energy balance for epitaxial growth of the 2D layer on graphene. To analyze this quantitatively, we employ a simple energy balance given by [33,34] (derivation in SI Section S4a)

$$E_{total} = -E_{vdW} - \Delta U_P\, f_{lock} + F_{strain}, \qquad (1)$$

where $E_{vdW}$ is the registry-independent van der Waals adhesion between *h*-CuI and graphene, $\Delta U_P$ is the corrugation amplitude, the additional binding gained when the two lattices reach an energetically favorable, commensurate stacking, and $f_{lock} \in [0, 1]$ is the lock-in factor (1 for biaxial commensurate registry, about 1/3 to 1/2 for uniaxial commensurate, 0 for the least favorable stacking; SI Section S4a). $E_{vdW}$ and $\Delta U_P$ are defined as positive quantities, so that the two leading minus signs identify them as the stabilizing contributions. $F_{strain}$, the elastic energy required to strain *h*-CuI toward commensurability, is positive and destabilizing; $E_{total}$ is referenced to the unstrained layer and the bare substrate (SI Section S4a). For complete commensurability, *h*-CuI would have to be stretched by the 1.7% lattice mismatch. In SI Section S4b we show that accommodating this biaxially would cost $F_{bi} \approx 1.05$ meV/Å$^2$, whereas a uniaxial match costs only $F_{uni} \approx 0.30$ meV/Å$^2$. The experimentally observed anisotropy of 1.4 – 1.7% over extended crystalline domains, see Table S1, shows that *h*-CuI adapts only uniaxially. The corrugation gain must therefore exceed the uniaxial cost but stay below the biaxial one, $0.30 < \Delta U_{P,eff} < 1.05$ meV/Å$^2$. The lower bound inferred from the observed uniaxial strain exceeds the corrugation amplitudes reported from DFT calculations in the literature (≤ 0.14 meV/Å$^2$, Table S3) [28] by about a factor of two, which we ascribe to the unavoidable simplification in the vdW potential assumed in those calculations.

Finally, we discuss how substrate defects and interfacial adsorbates affect the growth of *h*-CuI on graphene [29]. The balance of Eq. (1) holds for an idealized interface. DFT calculations for a defective graphene substrate (Methods) give an adhesion of 13.65 meV/Å$^2$, only 0.5% above the pristine value of 13.58 meV/Å$^2$; the single defective configuration examined therefore leaves the balance of Eq. (1) unchanged. In contrast, interfacial adlayers have a pronounced effect on the epitaxial growth. We estimate their effect with a planar potential for two parallel sheets [39] (SI Section S4d),

$$E_{vdW}(z) = E_0 \left[ (5/3)(z_0/z)^4 - (2/3)(z_0/z)^{10} \right], \quad (2)$$

where z is the graphene–CuI distance, $z_0$ = 3.68 Å is the equilibrium van der Waals gap (static DFT: 3.678 Å [26] and 3.675 Å [28]), and $E_0$ is the well depth, set to our calculated adhesion of 13.58 meV/Å$^2$; the $z^{-4}$ attractive tail is the pairwise dispersion limit for two parallel sheets [39]. An adlayer that increases the interfacial distance by Δz reduces the recovered adhesion to $E_{vdW}(z_0 + \Delta z)$. The thickness of a single interfacial adlayer is comparable to the vdW gap itself: adsorbed water layers between graphene measure 3.7 ± 0.2 Å [40] and hydration layers in liquid water 2.8–3.0 Å [41]. Inserting Δz = 2.8–3.7 Å into Eq. (2) leaves a recovered adhesion of only 1.4–2.3 meV/Å$^2$, that is 10–17% of $E_0$, which cannot compensate the 20.5 meV/Å$^2$ by which the 2D phase lies above bulk *γ*-CuI. Because such an adlayer is an equilibrium feature of the open graphene surface in liquid-phase synthesis, it offers an explanation for the absence of *h*-CuI on open monolayer regions in the wet-chemical route [26]. Within a bilayer, the van der Waals pressure expels part of the adsorbates into pockets [42,43] and presses the layers together, supporting the binding where adsorbates cannot be removed.

In our vapor-phase route the same balance appears in reverse: where adsorbates remain on the surface, the adhesion term of Eq. (1) is not recovered and growth stops. We compare two states of the interface, after HI treatment at 40 °C and after HI treatment followed by annealing at 180 °C. After the 40 °C HI treatment the reduced oxo-graphene surface still carries residual oxo-groups, physisorbed water, and adsorbed iodine [35], limiting the size of the *h*-CuI crystallites. At 180 °C, coalesced domains form (Fig. 3; Fig. 5), and STEM-EDX records a cleaner interface, with the oxygen signal dropping and sulfur and sodium falling below the detection limit (Fig. S1). This correlation is consistent with a contribution from adsorbate removal; the adsorbates at this interface desorb or migrate with barriers reported for other systems that span roughly 0.01–0.9 eV: oxo-group hopping on graphene 0.15–0.89 eV [36], copper edge diffusion 0.17 eV [37], and adsorbed iodine 10–20 meV [38]. The lateral domain size therefore correlates with the local cleanliness of the interface, as seen in Fig. S2, where the *h*-CuI crystals extend up to the next adsorbate on the surface. Where the interface is clean, further growth is expected to be limited mainly by precursor supply to the graphene–CuI interface.

**Stability of *h*-CuI on graphene**

In bulk, the layered hexagonal *β*-phase is an equilibrium phase only between 643 and 673 K and reverts to *γ*-CuI on cooling [16,21]. According to the observations, however, in two dimensions the phase persists to room temperature. Exfoliated flakes persist uncovered under ambient conditions [23]. Simulations find the freestanding lattice to be dynamically stable at room temperature [17,23]. A stabilizing effect of encapsulation in bilayer graphene was found in AIMD simulations, where the freestanding monolayer lost structural integrity at 600 K, while the encapsulated layer was still structurally stable [28]. Published calculations thus considered the freestanding monolayer [17,23] and the encapsulated layer [26,28] but not yet 2D *h*-CuI on an open graphene substrate, which is metastable by about 7.0 meV/Å$^2$ above bulk *γ*-CuI. To assess the stability of *h*-CuI also on SLG, we performed AIMD simulations at 300 and 600 K and compared the supported and encapsulated configurations. The cells used for the calculations were commensurate supercells with a typical twist angle of $\theta \approx 20.6°$ matching the annealed monolayer of Fig. 3(d), see SI Section S5. For each case at least six restart segments of about 1.5 ps were run, giving at least 9 ps of aggregate sampling per system (Section S5b).

We first analyze the collective movements within the *h*-CuI plane and relative to the additional graphene layers, Fig. 6(a)–(c). Panel (a) shows the power spectrum of the antisymmetric out-of-plane coordinate of the two CuI sublayers, $Z_{anti} = ½(\bar{z}_{top} - \bar{z}_{bot})$, at 300 K; see the inset of the figure for an illustration of the movement. Between 1.20 and 1.55 THz the oscillation in BLG-encapsulated and SLG-supported *h*-CuI shows a mean normalized power of 0.288 ± 0.074 and 0.028 ± 0.022, respectively (Table S5). We attribute the damping to the strong spectral coherence $\gamma^2$ between the out-of-plane center-of-mass motion of SLG and *h*-CuI, see Fig. 6(c). The curvature of the potential of Eq. (2) at its minimum defines a van der Waals spring normal to the plane [45,46], $k = 40E_0/z_0^2 \approx$ 40 meV/Å$^4$, or $6.4 \times 10^{19}$ N/m$^3$. On single-layer graphene such a spring force is present on one side of *h*-CuI while it is present on both sides in BLG. The areal masses of the CuI sheet and of one graphene layer are 25.9 and 4.6 amu/Å$^2$. Their reduced areal mass, $\mu$ = (25.9 × 4.6)/(25.9 + 4.6) = 3.9 in these units, places the layer-breathing mode of sheet against layer at $f = (1/2\pi)\sqrt{k/\mu}$ = 1.59 THz, just above the observed band at 1.20–1.55 THz. The same adhesion well depth used in the preceding section therefore also sets the expected frequency scale of this collective out-of-plane motion. The coherence of *h*-CuI with SLG exceeds that with either graphene layer of BLG by a factor of 2.5 to 6. Averaged over the analyzed range, 0.9 to 3.0 THz, $\gamma^2$ is 0.525 ± 0.050 on SLG, compared with 0.083 ± 0.134 and 0.206 ± 0.176 for the two graphene layers of BLG. Between 1.0 and 1.7 THz, $\gamma^2$ is 0.656 ± 0.067 on SLG, and 0.101 ± 0.317 and 0.070 ± 0.091 for the two BLG layers.

For the case of *h*-CuI on SLG, the sheet and the graphene layer move together in a common layer-breathing mode. Thus, a single graphene layer damps the relative out-of-plane movement of *h*-CuI between 1.20 and 1.55 THz by an order of magnitude in normalized power. The band near 4 THz is the internal Cu–I bond vibration, which is similar for both cases, on SLG and on BLG. Fig. 6(b) shows the power spectrum of the in-plane transverse optical coordinate, TO = $\bar{v}_y$(Cu) − $\bar{v}_y$(I), of the movement of the copper sublattice relative to the iodine sublattice. At 300 K, the TO bands are found at 1.93 ± 0.21 and at 2.76 ± 0.05 THz for the SLG and BLG cases, respectively. For this collective in-plane movement, a second interface stiffens the *h*-CuI sheet.

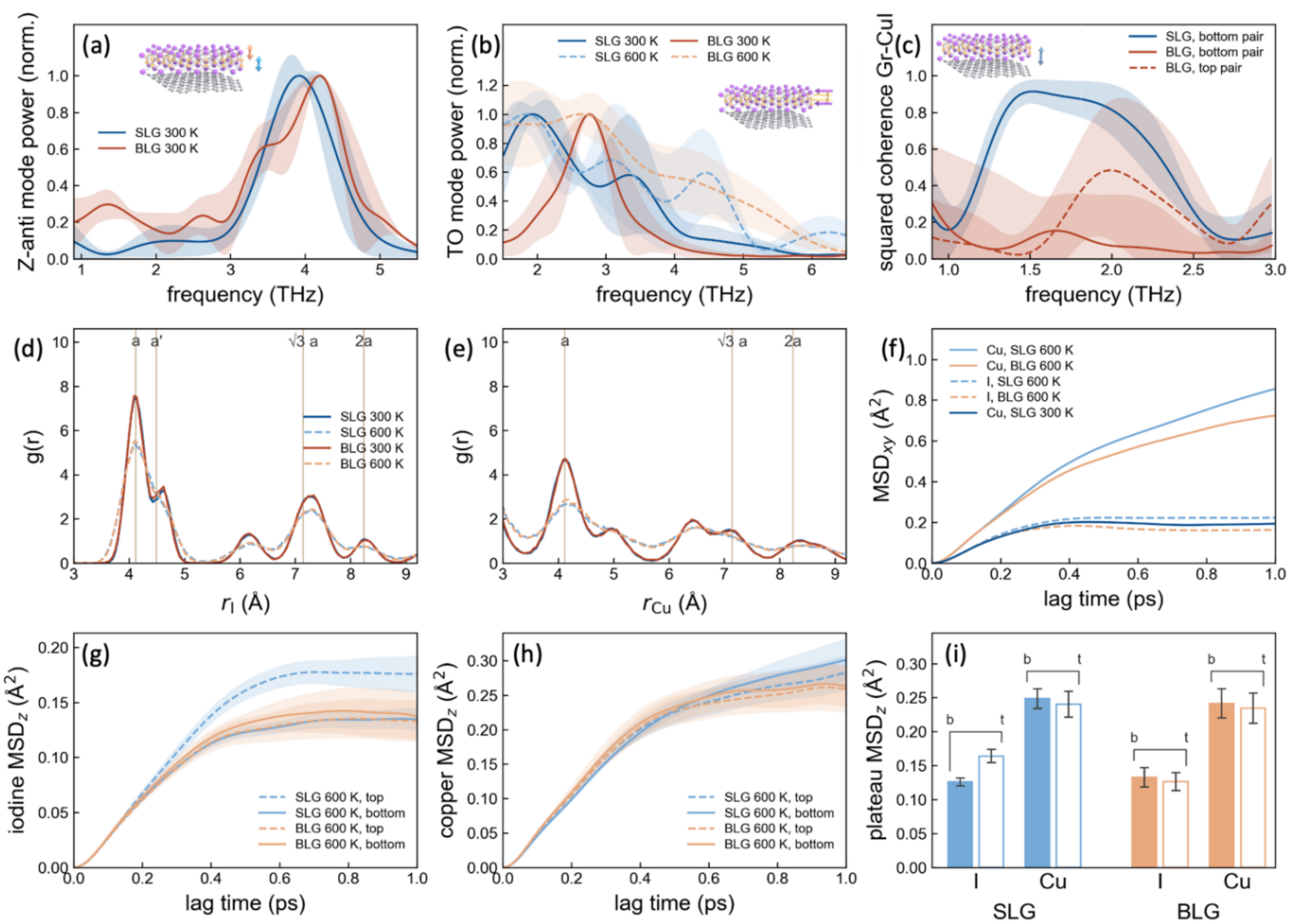

**Figure 6. Ab initio molecular dynamics simulations of *h*-CuI on single-layer graphene and between two graphene layers.** (a) Power spectrum of the antisymmetric out-of-plane coordinate of the two CuI sublayers, $Z_{anti} = ½(\bar{z}_{top} - \bar{z}_{bot})$, at 300 K; each curve is normalized to its own maximum. (b) Power spectrum of the in-plane transverse optical coordinate TO = $\bar{v}_y$(Cu) − $\bar{v}_y$(I) at 300 K (solid) and 600 K (dashed). (c) Squared spectral coherence $\gamma^2$ between the out-of-plane center-of-mass motion of a graphene layer and that of the adjacent CuI sublayer at 300 K. The insets in (a)–(c) illustrate the analyzed dynamics for the example of SLG. (d, e) Pair distribution functions g(r) of iodine, (d), and of copper, (e), at 300 and 600 K. Markers indicate the typical distances of the hexagonal lattice with a = 4.1 Å, √3a = 7.1 Å and 2a = 8.2 Å. a′ = 4.5 Å appears in the iodine panel for the nearest neighbor in the other iodine sublayer. (f) In-plane mean-square displacement $MSD_{xy}(\tau)$ at 600 K. (g, h) Out-of-plane mean-square displacement $MSD_z(\tau)$ of iodine and of copper at 600 K, for the two sublayers (bottom and top). In SLG the bottom sublayer faces graphene, see the illustration in the insets of (a)–(c). (i) Plateau values of (g) and (h), averaged over τ = 0.3–1.0 ps. The asymmetry is highlighted by a bracket on top of the bar chart; b and t denote the bottom and the top sublayer.

We then analyzed the pair distribution functions (PDF) g(r) of iodine with itself and of copper with itself inside the *h*-CuI to gain quantitative information on the structural stability. Fig. 6(d), (e) show the maxima at the characteristic distances of the double-hexagonal lattice with $a = 4.12$ Å, $\sqrt{3}a = 7.14$ Å and $2a = 8.24$ Å, where a is the neighbor distance within one sublayer. For iodine an additional maximum appears at $a' = 4.49$ Å corresponding to the nearest neighbor in the second iodine sublayer of the I–Cu–Cu–I sandwich. The characteristic maxima in the distribution functions show that the hexagonal lattice is retained at 300 K both on SLG and within BLG. At 600 K the iodine atoms remain strongly localized, while the Cu atoms show increased dynamics. At the first minimum of the Cu–Cu distribution, at 5.69 Å, g(r) rises from 0.11 to 0.77 on SLG and from 0.15 to 0.71 in BLG. The hexagonal order is still maintained at 600 K, mainly by the iodine framework, while the copper sublattice may delocalize inside it. Fig. 6(f) confirms that the in-plane mean-square displacement MSDxy(τ) at 600 K for Cu is much higher. The iodine displacement reaches a plateau of 0.220 Å$^2$, while the copper displacement keeps rising within the time frame of our simulation, see also Table S6. Over its longest continuous segment, 1.72 ps, a copper atom covers a mean net in-plane distance of 0.99 Å, a quarter of the neighbor distance. The copper therefore exhibits enhanced in-plane mobility within a framework maintained by the iodine sublattice. At 300 K, the movement of Cu within the I lattice is not observed; the $MSD_{xy}$ of Cu reaches the iodine plateau and then remains constant.

Finally, we analyze the consequence of the different type of clamping of *h*-CuI on SLG and BLG, respectively. Fig. 6(g)–(h) show the out-of-plane mean-square displacement $MSD_z(\tau)$ of iodine and of copper at 600 K. Each is resolved by sublayer, after removing the center-of-mass motion of that sublayer. For a bound atom the plateau equals $2\sigma_z^2$, so the amplitude follows as $\sigma_z = \sqrt{(MSD/2)}$. On SLG, the bottom sublayer faces graphene and the top faces vacuum; in the encapsulated cell both faces are in contact with graphene. The two iodine sublayers behave differently on graphene, reaching 0.126 Å$^2$ for the bottom layer and 0.164 Å$^2$ for the top layer, see Fig. 6(g). With graphene on both faces the sublayers stabilize at a similar $MSD_z$, 0.133 and 0.126 Å$^2$. The vdW spring of graphene thus damps the out-of-plane movement of the iodine sublayer in its vicinity. In BLG this damping is observed for both sublayers of *h*-CuI; in SLG the damping is asymmetric and stronger on the sublayer in direct contact with graphene. In Fig. 6(i) we summarize the plateau values of (g) and (h), averaged over τ = 0.3–1.0 ps, with b and t for the bottom and the top sublayer. The effect on Cu is markedly smaller, which we attribute to the larger distance of Cu compared to I from the van der Waals interface with graphene. The trajectories show that 2D *h*-CuI remains dynamically stable not only within a graphene sandwich but also on an open graphene surface at both 300 and 600 K.

# CONCLUSIONS

We have demonstrated the bottom-up growth of two-dimensional hexagonal copper(I) iodide (*h*-CuI) directly on reduced oxo-graphene. HI-vapor treatment at 40 °C initiates *h*-CuI nucleation, while annealing at 180 °C promotes the growth of extended crystalline domains on open monolayer graphene and within bilayer graphene. 80 kV Cc/Cs-aberration-corrected HRTEM resolves the atomic structure, local epitaxial alignment, and the characteristic uniaxial lattice anisotropy of the *h*-CuI domains, while STEM-EDX yields a Cu:I ratio consistent with 1:1. The lateral extent of the *h*-CuI domains correlates with the cleanliness of the interface and is limited mainly by remaining interfacial adsorbates.

DFT calculations give a van der Waals adhesion of 13.58 meV/Å² per graphene interface and identify *h*-CuI on open monolayer graphene as thermodynamically metastable, lying about 7.0 meV/Å² above bulk *γ*-CuI [21]. Under the present low-temperature precursor conditions, neither a molecular CuI vapor flux nor the local supersaturation required for *γ*-CuI nucleation is available, favoring selective formation of the hexagonal layer at the graphene interface. AIMD simulations further show that the hexagonal lattice remains intact at both 300 and 600 K on open monolayer graphene as well as within bilayer graphene. At 600 K, Cu atoms become increasingly mobile within an iodine framework that retains its hexagonal order.

The growth of oriented *h*-CuI on an open surface may be useful for its integration into heterostructure device platforms, combining the stability of graphene with the mechanical flexibility and favorable optoelectronic properties of the 2D metal halide. The predominantly van der Waals nature of the substrate interaction suggests that this approach is not restricted to graphene and *h*-CuI but may extend to other chemically inert 2D templates capable of stabilizing different metastable layered phases.

# EXPERIMENTAL METHODS

### Synthesis of *h*-CuI/r-oxo-G Heterostructures

Oxo-graphene (oxo-G) was synthesized using a modified Hummers method [47,48]. The oxo-G flakes were deposited onto standard copper TEM grids with an amorphous carbon support film (Quantifoil) by double self-assembly (DSA). In this process, oxo-G flakes self-assemble at the air–water interface and are compressed into a compact, non-overlapping film by sodium dodecyl sulfate (SDS) surfactant [49,50,52].

The oxo-G-coated copper grids were subsequently exposed to hydrogen iodide (HI) vapor to reduce the oxo-G template and initiate CuI formation. For this purpose, the samples were placed on a PTFE support inside a sealed reaction vessel containing 2 mL of aqueous HI solution (57 wt%). The reaction was carried out at 40 °C for 1 h. During this treatment, HI vapor acts both as a reducing agent for oxo-G, yielding reduced oxo-graphene (r-oxo-G), and as an iodine source for CuI formation. The copper TEM grid serves as a sacrificial solid-state copper precursor. Under these conditions, CuI nucleates on the graphene flakes.

To promote the growth of extended *h*-CuI domains, the samples were subsequently annealed at 180 °C for 2 h under ambient conditions. The essential role of the copper grid as the metal precursor was verified by control experiments using gold TEM grids, for which no *h*-CuI formation was observed under otherwise identical conditions. Further synthesis parameters and additional characterization are provided in the Supporting Information, Section S1.

**Materials Characterization**

Atomic-scale structural characterization was performed using the chromatic- and spherical-aberration- ($C_c$/$C_s$-) corrected Sub-Angstrom Low-Voltage Electron Microscope (SALVE), operated at an accelerating voltage of 80 kV to minimize beam-induced damage. The instrument employs a combined $C_c$/$C_s$ corrector that compensates the third-order spherical aberration $C_s$ and the off-axial coma $B_3$ together with the linear chromatic aberration, with residual geometric aberrations minimized up to fifth order [53]. Correcting $C_s$ removes the delocalization of lattice information, while the chromatic-aberration correction narrows the focus-spread envelope and thereby extends the information limit to 76 pm at 80 kV [53]. Measured values for $C_c$ and $C_s$ were in the range of −5 to −15 μm. The vacuum in the column of the TEM was ~$2\cdot10^{-7}$ mbar. Dose rates in the range of $10^5$ e/($nm^2\cdot s$) were used for the high-resolution images, and the images were recorded on a 4k × 4k CMOS camera with exposure times of 0.25–1 s. Chemical mapping and stoichiometric analysis were carried out by scanning transmission electron microscopy combined with energy-dispersive X-ray spectroscopy (STEM-EDX) using a Thermo Fisher Talos F200X operated at 80 kV and equipped with a high-sensitivity SuperX EDX detector.

The orientation and crystallinity of the *h*-CuI layers were analyzed from fast Fourier transforms (FFTs) of high-resolution TEM (HRTEM) images using the TemCompanion software suite [54]. Real-space *h*-CuI lattice vectors $b_1$, $b_2$, and $b_3$ were determined by fitting two-dimensional Gaussian functions to the atomic columns in the HRTEM images using Atomap with HyperSpy [31], enabling high-precision determination of local lattice parameters. The graphene primitive vectors $a_1$, $a_2$, and $a_3$ were extracted from the corresponding FFT reflections. All lattice measurements were internally calibrated against the graphene lattice constant $a_G$ = 2.460 Å, within the range of reported graphite basal-plane values of 2.4589 Å [55] to 2.4617 Å [56], which removes the pixel-size uncertainty common to all vectors. Details of the atomic-column fitting procedure, lattice-vector analysis, and twist-angle determination are provided in the Supporting Information, Sections S2a–S2c and Tables S1 and S2.

**Ab initio Calculations and Molecular Dynamics Simulations**

Static density functional theory (DFT) calculations of the interlayer binding between *h*-CuI and graphene were performed with the Vienna Ab initio Simulation Package (VASP) [57] within the projector augmented-wave (PAW) formalism [58]. The spin-polarized calculations used the Perdew-Burke-Ernzerhof (PBE) exchange–correlation functional [59], a plane-wave energy cut-off of 550 eV, and a Γ-centered Monkhorst–Pack k-point grid of 9 × 9 × 1. Dispersion was included through the DFT-D3 scheme [60] with Becke–Johnson damping [61]. Structures were relaxed until all residual forces were below 0.01 eV $Å^{-1}$, with an electronic convergence criterion of $10^{-6}$ eV.

The supercell used for the binding energy contained 18 Cu, 18 I, and 50 C atoms, that is, a 3 × 3 *h*-CuI cell on 5 × 5 graphene (rhombic cell of the hexagonal lattice, edge length 12.32 Å, area A = 131.4 $Å^2$). In this cell the *h*-CuI in-plane lattice constant is 4.107 Å, 2.0% below the freestanding value of 4.19 Å, so the layer is compressed rather than stretched; the (√3 × √3) commensuration discussed in the Vapor-phase growth on monolayer graphene section is a different registry. In this cell the *h*-CuI and the graphene hexagons are aligned, so the binding energies refer to the untwisted configuration and not to the twisted supercell of the molecular-dynamics runs described below. The binding energy was evaluated as $E_b$ = E(Gr) + E(*h*-CuI) − E(*h*-CuI@Gr), the three terms being the total energies of the graphene sheet, of the *h*-CuI layer, and of the heterostructure; with this sign convention $E_b$ is a positive adhesion energy. The adhesion energies quoted in the Growth mechanism and phase selection section, in the Stability of *h*-CuI on graphene section, and in the Supporting Information, Section S4c, are $E_b$ divided by the area of one interface. The same protocol was applied to pristine single-layer graphene and to defective graphene.

Ab initio molecular dynamics (AIMD) simulations were performed with the CP2K package [62] in the Gaussian plane-wave (GPW) scheme [63], again with the PBE functional [59]. The valence electrons were expanded in double-ζ valence basis sets with one set of polarization functions (DZVP) from the molecularly optimized family [64], while the core electrons and the nuclei were described by Goedecker–Teter–Hutter (GTH) pseudopotentials [65]. Four multi-grids with a plane-wave cut-off of 400 Ry were used, the Brillouin zone was sampled at the Γ point only, the nuclear propagation time step was 0.5 fs, and trajectories were run at 300 and 600 K. The molecular-dynamics supercell was larger than the binding-energy cell and was twisted. Two heterostructure configurations were considered within one commensurate supercell defined by the parameters $|R_1|$ = 24.99 Å, $|R_2|$ = 29.04 Å, γ = 37.5°, area $A_{cell}$ = 441.7 $Å^2$. The single-layer graphene (SLG) configuration consisted of one graphene layer and one *h*-CuI bilayer (60 Cu, 60 I, and 168 C atoms), whereas the bilayer graphene (BLG) configuration contained the same *h*-CuI bilayer encapsulated between two graphene layers (336 C atoms).

The integer commensurate matching, $R_1 = (6,0)_{h\text{-CuI}}$ versus $(9,2)_{Gr}$ and $R_2 = (3,5)_{h\text{-CuI}}$ versus $(3,10)_{Gr}$, corresponds to a lattice-vector angle of Δ ≈ 9.4°, i.e. an epitaxial twist of θ = |30° − Δ| ≈ 20.6° (Section S2c and Fig. S4), close to the twist of the experimentally annealed monolayer sample (θ = 21.2° ± 1.0°, Table S2). This construction reproduces the lattice constant of freestanding graphene within 0.21% ($a_G$ = 2.464 Å compared with the reported range of 2.4589–2.4617 Å). In *h*-CuI, the mean in-plane lattice constant is 4.157 Å, corresponding to a bidirectional compressive strain of −0.79% relative to the freestanding value of 4.19 Å.

The supercell geometry, the spectral quantities read from Figure 6, the statistics of the restart segments, and the mean-square-displacement analysis of the Stability of *h*-CuI on graphene section are provided in the Supporting Information, Section S5.

# ASSOCIATED CONTENT

**Supporting Information**

Additional details of the sample preparation, including double self-assembly deposition of oxo-graphene, hydrogen iodide vapor treatment, annealing, and the control synthesis on a gold grid; characterization methods, including determination of the *h*-CuI and graphene lattice vectors, twist-angle determination, and additional STEM-EDX data; additional bilayer graphene samples; energy balance calculation; analysis and methods description of the AIMD simulations; Figures S1–S5 and Tables S1–S7 (PDF)

# ACKNOWLEDGMENTS

D.K., A.T., G.J., J.P. and B.D.I. acknowledge the financial support from Thuringian Ministry of Economic Affairs, Science and Digital Society (TMWWDG) in Germany within the projects ESTI (2019 FGR 0080) and GraphSens (2020 FGR 0051) co-financed by the European Union (EU) within the European Social Fund (ESF). J.K., U.K., and A.N.K. acknowledge the funding of the Deutsche Forschungsgemeinschaft in the frame of the project DELAVER, funding Number 471707562. S.G. and E.B. acknowledge the financial support by the EPSRC Programme Grant 'Metal Atoms on Surfaces and Interfaces (MASI) for Sustainable Future' (EP/V000055/1). CPU time is provided by the ARCHER2 UK National Supercomputing Service, and the Sulis Tier 2 HPC platform funded by EPSRC Grant EP/T022108/1 and the HPC Midlands+ consortium. C.E.H. and S.E. gratefully acknowledge funding by the Deutsche Forschungsgemeinschaft (DFG, German Research Foundation) through CRC 1772 (C01, project ID 555467911). We thank Michael Seifert from the Friedrich Schiller University Jena, Germany, for fruitful discussions. AI-based language tools were used to assist in editing and formatting the manuscript text; all scientific content, analyses, and conclusions were developed and verified by the authors, who take full responsibility for the published content.

# AUTHOR CONTRIBUTIONS

J.K. performed the HRTEM measurements, and J.K. and D.K. analyzed the HRTEM data. G.J. prepared the oxo-graphene deposition and performed the vapor-phase synthesis. S.G. performed the DFT and AIMD simulations, and D.K. and S.G. analyzed the simulation data. A.T. initiated and supervised the collaboration regarding the oxo-graphene templates and sample preparation. C.E.H. and S.E. synthesized the oxo-graphene material. B.D.-I. and J.P. supervised the *h*-CuI synthesis. E.B. supervised the first-principles calculations. A.K. contributed to the interpretation of the first-principles results and to the discussion. U.K. supervised the TEM studies and contributed to the discussion. A.T., B.D.-I., E.B., A.K. and U.K. acquired funding for the project. The manuscript was written by D.K., G.J., U.K., S.G., and E.B. with contributions of all co-authors.

## NOTES

The authors declare no competing financial interest.

# DATA AVAILABILITY

The data supporting the findings of this study are available from the corresponding author upon reasonable request.

# Supporting Information

## S1. Sample Preparation

The synthesis of nanohybrid heterostructures of *h*-CuI on reduced oxo-graphene [S1] follows the steps (A)–(F) listed below. In brief, oxo-graphene (oxo-G) [S2] is deposited directly onto a copper TEM grid with an amorphous carbon support film (Quantifoil) using the Double Self-Assembly (DSA) method [S3]. Subsequently, the sample is exposed to hydrogen iodide (HI) vapor. This treatment reduces the oxo-G and simultaneously initiates the nucleation of small *h*-CuI crystallites, formed from iodine supplied by the reducing agent and copper provided by the TEM grid. Finally, the sample undergoes thermal annealing at 180 °C, which promotes the reaction between iodine and copper to form extended, well-defined *h*-CuI crystals on the reduced oxo-G (r-oxo-G) substrate or sandwiched between the graphene nanosheets of bilayer r-oxo-G.

**A) Materials and Substrate Preparation:** A suspension of oxo-G flakes in a 1:1 volume ratio of deionized water and isopropanol was used as the initial material. Standard copper TEM grids (Quantifoil with amorphous carbon support) were used as substrates. Prior to deposition, the TEM grids and the petri dishes used for the assembly were thoroughly cleaned. They were placed in an ultrasonic bath with acetone for 15 minutes, followed by multiple rinses with high-purity water (18.2 MΩ·cm), and finally blow-dried with nitrogen gas.

**B) Film Assembly at the Air-Water Interface:** The cleaned petri dish was filled with ultrapure water. The oxo-G suspension, which was homogenized for 30 minutes in an ultrasonic bath, was then carefully dispensed onto the water's surface using a syringe until the entire surface was covered with a loosely arranged, floating layer of oxo-G flakes.

**C) Film Compression:** To compress this layer, a 10% aqueous solution of sodium dodecyl sulfate (SDS) was introduced dropwise from a separate syringe at the edge of the petri dish. The SDS molecules rapidly spread across the water's surface, acting as a mobile barrier that pushed the floating oxo-G flakes to the opposite side of the dish. This process caused the flakes to form a stable, compact, and uniform monolayer film. The compression was analogous to the Langmuir–Blodgett process, with the difference that the SDS molecules rearranged by self-assembly, so that a compact, uniform monolayer was maintained throughout the deposition.

**D) Transfer onto TEM Grids:** The liquid was carefully drained from the petri dish, leaving the compact oxo-G film floating on a thin layer of residual water. The film was then captured by carefully bringing the TEM grid up from underneath the film, allowing the film to settle and adhere to the grid's surface. Finally, the grid with the deposited film was heated in an oven at 80 °C for 15 minutes to ensure it was thoroughly dried before subsequent synthesis steps.

**E) HI-Vapor Treatment of oxo-G Films:** The HI-vapor treatment was performed in a sealed glass container equipped with a Teflon sample support. First, 2 ml of an aqueous hydrogen iodide solution (57 wt%) was placed at the bottom of the container. The oxo-graphene-coated TEM grid was then positioned on the Teflon support, ensuring it was suspended above the liquid and not in direct contact with it. The container was sealed and placed in a furnace, where it was heated to 40 °C for one hour.

**F) Annealing:** Following the HI treatment, the samples were annealed in a separate furnace at 180 °C for two hours under ambient conditions.

To verify that the copper TEM grid serves as the essential copper source for *h*-CuI formation, an identical synthesis was performed using a gold (Au) TEM grid instead. The oxo-G film was deposited onto the Au grid following the same DSA protocol, subjected to the same HI vapor treatment at 40 °C for one hour, and annealed at 180 °C for two hours. Under these conditions, no *h*-CuI crystals were observed. The absence of *h*-CuI formation on the Au grid confirms that the copper from the Cu TEM grid is the sole source of Cu atoms in our synthesis, validating the solid-state precursor approach. In HAADF-STEM images of HI-treated and annealed samples, bright high-contrast particles are observed mainly on the holey-carbon support; EDX identifies them as residual gold contamination.

## S2. Characterization Methods

Scanning transmission electron microscopy (STEM) and energy-dispersive X-ray spectroscopy (EDX) were performed on a Thermo Fisher Talos F200X (S)TEM operated at an acceleration voltage of 80 kV. High-resolution atomic structure imaging was conducted using the $C_c/C_s$-corrected "Sub-Angstrom Low-Voltage Electron Microscope" (SALVE) instrument at 80 kV [S4].

### S2a. *h*-CuI Lattice Vectors

High-resolution TEM images were analyzed using Atomap with HyperSpy [S5]. Atomic column positions were identified using peak finding and refined by fitting 2D Gaussian functions to each atomic column, achieving a precision of ~1 pm. The 2D Gaussian fit provides the position (x, y), amplitude, and shape parameters ($\sigma_x$, $\sigma_y$, rotation angle) for each atomic column. From the detected atom positions, three principal lattice directions ($\mathbf{b}_1$, $\mathbf{b}_2$, $\mathbf{b}_3$) are automatically identified, corresponding to the three symmetry-equivalent translation directions of the hexagonal lattice, each separated by ~60°; for Table S1 they are relabeled in ascending order of length. The graphene lattice constant ($a_G$ = 2.460 Å = 0.2460 nm; Methods of the main text) served as internal calibration for the nm/pixel in the HRTEM image.

**S2b. Graphene Lattice Vectors**

FFT performed using TemCompanion software [S6] was used to determine the angle of the graphene lattice vectors. The same internal calibration ($a_G$) as in S2a was applied.

**S2c. Twist Angle Determination**

The epitaxial twist angle θ = |30° − Δ| quantifies the deviation from perfect 30° alignment between *h*-CuI and graphene, where Δ is the angular difference between the nearest *h*-CuI and graphene lattice vectors (Table S2).

The direct-FFT inter-graphene twist $\phi_{GG'}$ = 21.7 ± 1.0° is reported for the annealed bilayer crystallites that align to a single graphene layer, Sample 4 in Fig. 4(d–f) (*h*-CuI aligned with G') and Sample 5 in Fig. S3 (aligned with G). The ±1.0° uncertainty reflects that the inter-layer twist cannot be read from the diffraction spots to better than about one degree. For these two samples the *h*-CuI lattice is aligned with one graphene layer at the commensurate 30° orientation, so the entries for that layer are quoted as idealized values (Δ' ≈ 30° and θ' ≈ 0° for Sample 4; θ ≈ 0° for Sample 5). The angles to the other layer follow from the inter-graphene twist as Δ = 8.3° (Sample 4) and Δ' = 8.3° (Sample 5), i.e. 30° − 21.7°, with θ = θ' = 21.7°; these values are average values of the vectors calculated from atomap, 7.7° and 9.1°, for the two samples.

**Table S1.** Lattice vectors for *h*-CuI/graphene heterostructures. Samples 1–2 = monolayer graphene (SLG); Samples 3–5 = bilayer graphene (BLG). $a_1$, $a_2$, $a_3$ = graphene (G) lattice vectors from FFT; $a_1'$, $a_2'$, $a_3'$ = lattice vectors of the second graphene layer (G'); $b_1$, $b_2$, $b_3$ = *h*-CuI lattice vectors from Atomap; $A_b$ = anisotropy. T=40 °C: HI vapor treatment at 40 °C only, T=40/180 °C: HI vapor treatment at 40 °C followed by thermal annealing at 180 °C. The *h*-CuI vectors are labeled in ascending order of length; their magnitudes are computed from the unrounded components. The graphene lattice constant $a_G$ = 0.2460 nm serves as internal calibration (Section S2a), |a|=0.246 nm for all graphene lattice vectors.

| Sample | Substrate / Growth temperature | Graphene Lattice vectors | *h*-CuI Lattice vectors |
|---|---|---|---|
| 1 (see Fig. 3) | r-oxo-G monolayer / 40 °C | $a_1$=(0.108, −0.221) nm<br>$a_2$=(0.138, 0.204) nm<br>$a_3$=(0.245, −0.017) nm | $b_1$=(0.060, −0.425) nm<br>$\|b_1\|$=0.429 ± 0.002 nm |
| | | | $b_2$=(0.400, −0.162) nm<br>$\|b_2\|$=0.431 ± 0.002 nm |
| | | | $b_3$=(−0.340, −0.264) nm<br>$\|b_3\|$=0.431 ± 0.002 nm |
| | | | $A_b$=1.005 ± 0.004 |
| 2 (see Fig. 3) | r-oxo-G monolayer / 40 °C /180 °C | $a_1$=(0.170, 0.178) nm<br>$a_2$=(0.239, −0.058) nm<br>$a_3$=(−0.069, 0.236) nm | $b_1$=(0.056, −0.418) nm<br>$\|b_1\|$=0.421 ± 0.002 nm |
| | | | $b_2$=(0.395, −0.158) nm<br>$\|b_2\|$=0.425 ± 0.002 nm |
| | | | $b_3$=(−0.339, −0.260) nm<br>$\|b_3\|$=0.427 ± 0.002 nm |
| | | | $A_b$=1.014 ± 0.004 |
| 3 (see Fig. 4) | r-oxo-G bilayer / 40 °C | $a_1$=(−0.003, 0.246) nm<br>$a_2$=(0.214, −0.121) nm<br>$a_3$=(0.212, 0.125) nm<br>$a_1'$=(0.110, −0.220) nm<br>$a_2'$=(0.246, −0.015) nm<br>$a_3'$=(0.136, 0.205) nm | $b_1$=(−0.291, −0.298) nm<br>$\|b_1\|$=0.416 ± 0.002 nm |
| | | | $b_2$=(0.116, −0.399) nm<br>$\|b_2\|$=0.416 ± 0.002 nm |
| | | | $b_3$=(0.407, −0.101) nm<br>$\|b_3\|$=0.419 ± 0.002 nm |
| | | | $A_b$=1.007 ± 0.004 |
| 4 (see Fig. 4) | r-oxo-G bilayer / 40 °C /180 °C | $a_1$=(0.057, −0.239) nm<br>$a_2$=(0.236, −0.070) nm<br>$a_3$=(−0.178, −0.169) nm<br>$a_1'$=(−0.245, −0.026) nm<br>$a_2'$=(0.145, −0.199) nm<br>$a_3'$=(−0.100, −0.225) nm | $b_1$=(0.041, −0.417) nm<br>$\|b_1\|$=0.419 ± 0.002 nm |
| | | | $b_2$=(0.387, −0.168) nm<br>$\|b_2\|$=0.422 ± 0.002 nm |
| | | | $b_3$=(−0.345, −0.249) nm<br>$\|b_3\|$=0.426 ± 0.002 nm |
| | | | $A_b$=1.017 ± 0.004 |

| **Sample** | **Substrate / Growth temperature** | **Graphene Lattice vectors** | ***h*-CuI Lattice vectors** |
|---|---|---|---|
| 5 (see Fig. S3) | r-oxo-G bilayer / 40 °C /180 °C | $a_1$=(0.178, 0.169) nm<br>$a_2$=(0.057, −0.239) nm<br>$a_3$=(0.236, −0.070) nm<br>$a_1'$=(0.145, −0.199) nm<br>$a_2'$=(0.100, 0.225) nm<br>$a_3'$=(0.245, 0.026) nm | $b_1$=(0.297, −0.294) nm<br>$\lvert b_1\rvert$=0.418 ± 0.002 nm |
| | | | $b_2$=(−0.114, −0.405) nm<br>$\lvert b_2\rvert$=0.421 ± 0.002 nm |
| | | | $b_3$=(0.411, 0.111) nm<br>$\lvert b_3\rvert$=0.426 ± 0.002 nm |
| | | | $A_b$=1.019 ± 0.004 |

**Table S2.** Epitaxial twist angle calculation. $b_3$ = orientation of the *h*-CuI lattice vector $b_3$ (Table S1); a = nearest graphene (G) lattice vector orientation; Δ = angle between lattice vector of graphene layer G and *h*-CuI; Δ' = angle between lattice vector of graphene layer G' and *h*-CuI; θ = |30° − Δ| = deviation from 30° perfect lock-in. Orientations are given modulo 180°.

| Sample | $b_3$ (°) | a (°) | Δ (°) | θ (°) | Δ' (°) | θ' (°) |
|---|---|---|---|---|---|---|
| 1 | 37.8 ± 1.0 | 55.9 ± 1 | 18.1 ± 1.0 | 11.9 ± 1.0 | — | — |
| 2 | 37.5 ± 1.0 | 46.3 ± 1 | 8.8 ± 1.0 | 21.2 ± 1.0 | — | — |
| 3 | −13.9 ± 1.0 | −29.5 ± 1 | 15.6 ± 1.0 | 14.4 ± 1.0 | 10.4 ± 1.0 | 19.6 ± 1.0 |
| 4 | 35.8 ± 1.0 | 43.5 ± 1 | 8.3 ± 1.0 | 21.7 ± 1.0 | ≈ 30 | ≈ 0 |
| 5 | 15.1 ± 1.0 | 43.5 ± 1 | 28.4 ± 1.0 | ≈ 0 | 8.3 ± 1.0 | 21.7 ± 1.0 |

### S2d. Additional STEM-EDX Characterization

Fig. S1 presents the individual EDX elemental maps of the composite map shown in Fig. 5 of the main text.

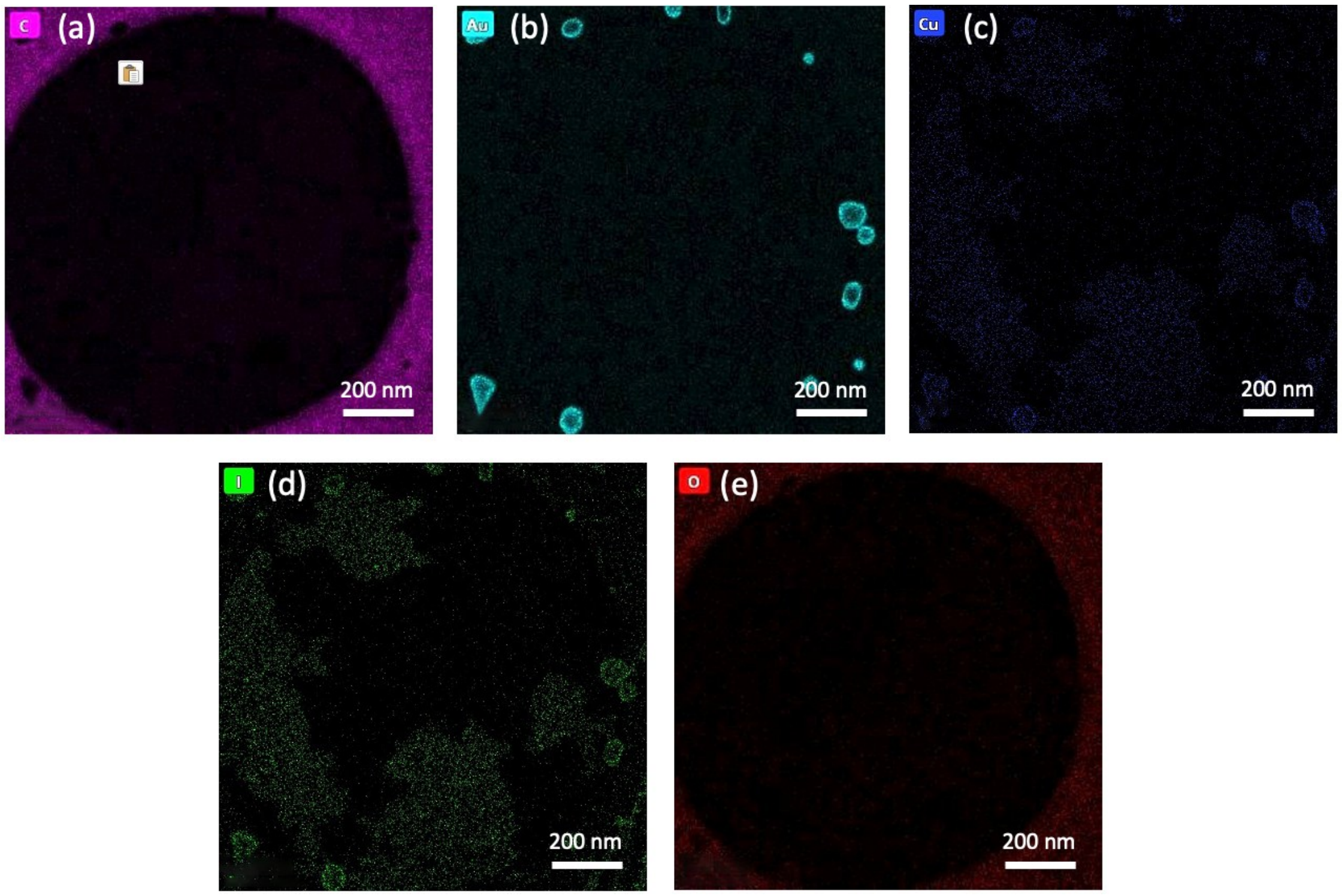

**Figure S1. Individual EDX elemental maps corresponding to the STEM-EDX analysis in Fig. 5 of the main text.** (a) Composite EDX map showing the spatial distribution of all detected elements. (b) Carbon EDX map (C-K). (c) Copper EDX map (Cu-L). (d) Iodine EDX map (I-L). (e) Oxygen EDX map (O-K), showing reduced oxygen content in the crystal regions after annealing.

## S3. Additional Bilayer Graphene Samples

Fig. S2 resolves individual *h*-CuI crystallites at atomic resolution after annealing. Two adjacent domains, false-colored cyan and magenta, each show the continuous hexagonal *h*-CuI lattice, while the surrounding graphene is contaminated. The lateral domain sizes are about 20 and 27 nm. The domains terminate where they meet these contaminated regions, consistent with the contamination-limited growth identified in the Formation Mechanism and Stability section of the main text. The two domains are the crystallites analysed in Fig. 4(d–f) (Sample 4) and in Fig. S3 (Sample 5); the panels of both figures are taken from this field of view. Where they meet, they abut along a common boundary without overlapping, so both grew in the same plane.

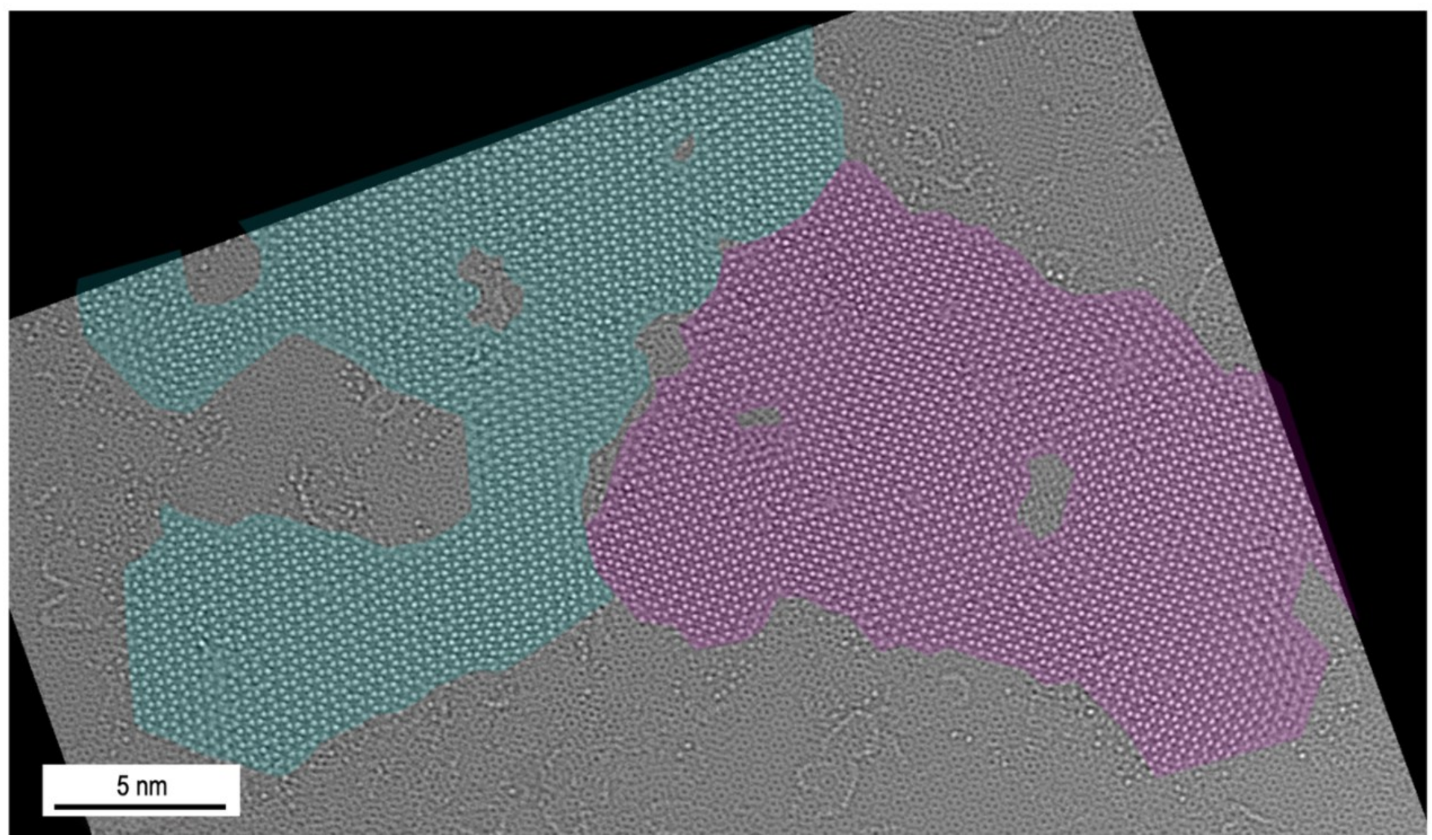

**Figure S2. Atomic-resolution HRTEM of *h*-CuI crystallites within reduced oxo-graphene bilayer after annealing.** Two adjacent crystalline domains (false-colored cyan and magenta), with lateral sizes of about 20 and 27 nm against the 5 nm scale bar (within the up to about 30 nm range reported in the main text), terminate at the surrounding amorphous, contaminated regions.

Both crystallites are located within the same bilayer flake: Table S1 lists the same six graphene lattice vectors for Sample 4 and for Sample 5, so G and G' denote the same two sheets in both cases. Sample 4 is nevertheless locked at the commensurate 30° orientation to G' and Sample 5 to G, each keeping 21.7 ± 1.0° to the respective other sheet (Table S2), and Fig. S2 shows the two domains meeting in one plane. A crystallite adsorbed on the outer surface would be templated by the outermost sheet, which is the same sheet for both domains. Registry with different sheets from within one plane is available only in the gallery, which places the bilayer crystallites between the graphene layers.

Fig. S3 shows the additional *h*-CuI crystal within bilayer graphene after annealing (Sample 5 in Table S1). The FFT inset in panel (a) confirms the bilayer substrate through the presence of two rotated graphene lattices (G, G'). The FFT of the heterostructure region [Fig. S3(b)] reveals that the *h*-CuI diffraction spots (red dashed hexagon) nearly overlap with those of G, indicating $\theta \approx 0°$ (Table S2). The crystal therefore adopts epitaxial alignment with G rather than with the second graphene layer G'. The structure model [Fig. S3(c)], reconstructed from the experimental lattice vectors and orientation mapping as in Figs. 3 and 4, shows this reversed alignment configuration in top and side view. This crystal complements the bilayer sample documented in Fig. 4 of the main text. Together they show that *h*-CuI can adopt either graphene layer as its epitaxial template.

# S4. Energy Balance for *h*-CuI Growth on Graphene

## S4a. Model

The moiré total-energy density of a 2D layer on a misoriented substrate reads (Woods 2014 [S7], based on the Frank–van der Merwe misfit theory [S8]):

$$E_{total} = \langle V(\mathbf{r} + \mathbf{u}(\mathbf{r})) \rangle_{AM} + (1)/(2)\langle C_{ijkl}\, \varepsilon_{ij}(\mathbf{r})\, \varepsilon_{kl}(\mathbf{r}) \rangle_{AM} \quad \text{(S1)}$$

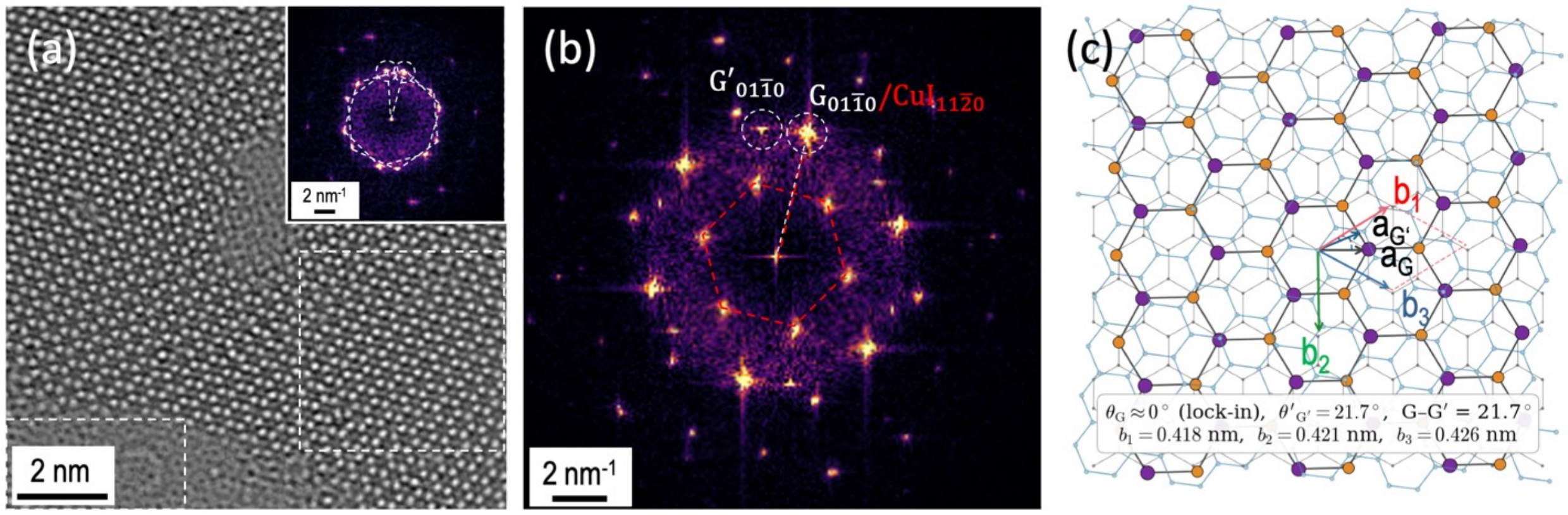

**Figure S3. Additional *h*-CuI/BLG(ann.) crystal with reversed epitaxial alignment (Sample 5).** (a) $C_c/C_s$-corrected 80 kV HRTEM image of an *h*-CuI crystal within bilayer graphene after annealing; inset: FFT of the substrate region confirming bilayer nature (G, G'). (b) FFT of the heterostructure region showing that *h*-CuI (red dashed hexagon) aligns with G at a 30° rotation (θ ≈ 0°), while G' appears as a separate hexagonal pattern. (c) Structure model of *h*-CuI encapsulated within bilayer graphene, shown in top and side view.

where $\langle ... \rangle_{AM}$ is the moiré-cell average, V the registry potential, u(r) the in-plane relaxation field, and $C_{ijkl}$ the 2D-elastic tensor. Assuming a rigid layer (uniform strain $\varepsilon_{ij}$, no in-cell relaxation $u(r) = u_0$), the registry average reduces to a discrete lock-in approximation:

$$\langle V(\mathbf{r} + \mathbf{u}_0)\rangle \approx -E_{vdW} - \Delta U_P\, f_{lock} \quad \text{(S2)}$$

with $f_{lock} \in [0, 1/3]$ for incommensurate registry, [1/3, 1/2] for uniaxial commensurate, and 1 for biaxial commensurate (the bounds reflect the Pokrovsky–Talapov continuum result). With the 2D elastic strain energy $F_{strain}$ (Section S4b), we obtain:

$$E_{total} = -E_{vdW} - \Delta U_P \cdot f_{lock} + F_{strain} \quad \text{(S3)}$$

where $E_{vdW}$ is the registry-independent van der Waals adhesion (positive, stabilizing), $\Delta U_P \cdot f_{lock}$ is the corrugation gain from commensurate lock-in ($f_{lock}$ as defined above), and $F_{strain}$ is the elastic strain energy (positive, destabilizing).

## S4b. Strain Energy Formulas

For the general 2D strain (experimental data with full tensor) [S9]:

$$F = (E_{2D})/(2(1 - v^2))[ \varepsilon_{xx}^2 + \varepsilon_{yy}^2 + 2v\, \varepsilon_{xx}\varepsilon_{yy} + 2(1 - v)\, \varepsilon_{xy}^2 ] \quad \text{(S4)}$$

For biaxial strain ($\varepsilon_{xx} = \varepsilon_{yy} = \varepsilon$, $\varepsilon_{xy} = 0$):

$$F_{bi} = (E_{2D})/(1 - v)\, \varepsilon^2 \quad \text{(S5)}$$

For uniaxial strain along one lattice direction ($\varepsilon_{xx} = \varepsilon$, $\varepsilon_{yy} = 0$, $\varepsilon_{xy} = 0$), the laterally constrained upper estimate:

$$F_{uni}^{\varepsilon} = (E_{2D})/(2( 1 - v^2 ))\, \varepsilon^2 \quad \text{(S6)}$$

This laterally constrained form ($\varepsilon_{yy} \approx 0$) is the upper estimate of the uniaxial strain cost, applicable when the stiffer graphene substrate ($E_{2D}^{Gr} \approx 340$ N/m) suppresses lateral relaxation of the soft *h*-CuI ($E_{2D} \approx 35$ N/m, Table S3). In Eqs. (S6) and (S7) the superscript names the transverse quantity that is held fixed: ε for zero transverse strain, where the substrate constrains the layer laterally, and σ for zero transverse stress, where the layer is free to contract laterally by the Poisson amount.

For the laterally free case with full Poisson relaxation (uniaxial-stress limit, $\varepsilon_{xx} = \varepsilon$, $\varepsilon_{yy} = -v\varepsilon$, $\varepsilon_{xy} = 0$), consistent with the experimentally observed in-plane contraction ($b_1 < a_{CuI} = 4.19$ Å for Samples 3 and 5, Table S1):

$$F_{uni}^{\sigma} = (1)/(2)\; E_{2D}\, \varepsilon^2 \quad \text{(S7)}$$

The strain anisotropy ratio is:

$$(F_{bi})/(F_{uni}^{\sigma}) = (2)/(1 - v) = 3.45 \quad \text{(S8)}$$

This anisotropy factor [Eq. (S8)] helps explain the preference for uniaxial adaptation observed experimentally.

The experimental in-plane lattice mismatch is $\delta \approx 1.67\%$ ($a_{CuI} = 4.19$ Å, $\sqrt{3}\cdot a_G = 4.26$ Å, see Table S3), and the layer must take it up as a strain $\varepsilon = \delta$. For this mismatch the three strain configurations above predict the following *h*-CuI lattice-vector lengths ($b_1$, $b_2$, $b_3$ labeled in ascending order of length as in Table S1, with the strain axis along the largest vector; the two shorter vectors are at 60° to that axis and therefore lengthen by about a quarter of the applied strain even when the transverse strain vanishes) and anisotropy $A_b = \max(b)/\min(b)$:

- Uniaxial, laterally constrained (no Poisson relaxation, Eq. (S6)): $b_1 = 4.208$ Å, $b_2 = 4.208$ Å, $b_3 = 4.260$ Å, $A_b = 1.012$.

• Uniaxial, laterally free (full Poisson relaxation, Eq. (S7)): $b_1$ = 4.186 Å, $b_2$ = 4.186 Å, $b_3$ = 4.260 Å, $A_b$ = 1.018.

• Biaxial commensurate (stretched): $b_1 = b_2 = b_3$ = 4.260 Å, $A_b$ = 1.000.

The uniaxial cost depends on how freely the monolayer contracts perpendicular to the relaxation direction and lies between the Poisson-relaxed value, $F_{uni}^{\sigma} \approx 0.30$ meV/Å$^2$ (Eq. (S7)), and the laterally constrained value, 0.37 meV/Å$^2$ (Eq. (S6)). The ratio of the two costs is $F_{bi}/F_{uni}^{\sigma} = 2/(1-\nu) = 3.45$. The main text (Growth mechanism and phase selection) estimates limits for the effective corrugation between the lower uniaxial limit and the biaxial upper limit, $0.30 < \Delta U_{P,eff} < 1.05$ meV/Å$^2$; this lies above the rigid-sliding upper bound of Table S3. The measured lattice anisotropy of the two annealed, graphene-aligned crystallites, 1.017 ± 0.004 (Sample 4) and 1.019 ± 0.004 (Sample 5, Table S1), matches the laterally free prediction of 1.018 rather than the laterally constrained value of 1.012, so the uniaxial cost is quoted at the lower limit. The contraction seen in Sample 5 ($b \approx 4.18$ Å $< a_{CuI} = 4.19$ Å, Table S1) confirms the perpendicular Poisson contraction.

**S4c. Van der Waals Adhesion: Direct DFT and Literature Cross-Check**

The primary adhesion values used in the main text were obtained by direct DFT in the untwisted 3 × 3 *h*-CuI on 5 × 5 graphene supercell, see the Methods section of the main text (VASP, PBE+D3 with Becke–Johnson damping; 18 Cu, 18 I and 50 C atoms; rhombic cell, A = 131.4 Å$^2$). The adhesion was evaluated as $E_{vdW}$ = [E(Gr) + E(*h*-CuI) − E(hetero)] / A. The calculated values are 13.58 meV/Å$^2$ (0.218 J/m$^2$) for pristine graphene and 13.65 meV/Å$^2$ for defective graphene.

Note on the $\Delta U_P$ uncertainty in Table S3. The Gr–CuI corrugation potential $\Delta U_P$ is reported as 0.01–0.14 meV/Å$^2$ (central value 0.07). The upper bound of this range is derived from Ullah et al. [S11], who computed the registry-dependent stacking energy of CuI encapsulated by graphene by sliding and rotating the layers across five distinct configurations and found the total energy variation to be < 0.3 meV/atom (DFT, vdW-DF1/optPBE-vdW). Their supercell contains 14 atoms (12 graphene C + 2 CuI atoms) over an area $A = (\sqrt{3}/2)\cdot a_{CuI}^2 \approx 15.20$ Å$^2$ with two equivalent CuI/graphene interfaces, yielding a total per-area span ≤ 0.3 × 14 / 15.20 ≈ 0.276 meV/Å$^2$ for the encapsulated cell, or ≤ 0.14 meV/Å$^2$ per single CuI/graphene interface as relevant for the one-sided geometry studied here. This per-interface DFT span sets the upper end of the quoted range (0.14 meV/Å$^2$); the lower end of the range (0.01 meV/Å$^2$) is not obtained by subtraction but represents the physical floor of the vanishing-corrugation regime consistent with the experimentally observed soft twist-angle degree of freedom. The same uncertainty range propagates to all derived quantities involving $\Delta U_P$.

A sensitivity analysis of the γ → β phase penalty covering the literature-bulk range and a conservative 2×-bulk upper bound is included in Table S4. The freestanding open monolayer lies about 20.5 meV/Å$^2$ above bulk γ-CuI, 17.4 meV/Å$^2$ for the CuI–CuI interlayer bond. One graphene interface (13.58 meV/Å$^2$) reduces the energy difference of *h*-CuI and bulk γ-CuI, however leaving the *h*-CuI/SLG heterostructure still about +7.0 meV/Å$^2$ above γ (metastable, kinetically trapped); a second interface further reduces the energy difference to −6.6 meV/Å$^2$ (thermodynamically favored).

### S4d. Adhesion in the Presence of an Interfacial Gap

To evaluate the effect of interfacial adlayers, the sheet–sheet interaction is modeled with the planar potential [S14], $E_{vdW}(z) = E_0 [(5/3)(z_0/z)^4 - (2/3)(z_0/z)^{10}]$. This form is calibrated to the direct-DFT depth $E_0$ = 13.58 meV/Å$^2$ at the equilibrium gap $z_0$ = 3.68 Å. Its curvature at $z_0$ gives the interfacial van der Waals spring constant, $k = 40\ E_0/z_0^2 \approx 40$ meV/Å$^4$ = 6.4 × 10$^{19}$ N/m$^3$. With the areal mass of the *h*-CuI sheet, 25.9 amu/Å$^2$, and that of one graphene layer, 4.6 amu/Å$^2$, the reduced areal mass is 3.9 amu/Å$^2$, and the interfacial breathing mode follows as $f = (1/2\pi)\sqrt{(k/\mu)}$ = 1.59 THz. The adhesion decays steeply due to the $z^{-4}$ tail. An interfacial layer, such as the adsorbed water layer, 3.7 ± 0.2 Å thick [S15], that is ubiquitous on graphitic surfaces under ambient conditions [S16], is comparable to $z_0$ itself. Evaluating $E_{vdW}(z_0 + \Delta z)$ for an additional gap $\Delta z$ = 2.8–3.7 Å gives 2.3 meV/Å$^2$ at $\Delta z$ = 2.8 Å and 1.4 meV/Å$^2$ at $\Delta z$ = 3.7 Å, that is 17% and 10% of $E_0$. Such an interfacial layer therefore strongly reduces the van der Waals stabilization required for growth of supported *h*-CuI.

**Table S3.** Lattice and DFT-input parameters used in the energy balance calculation.

| Parameter | Symbol | Value | Unit | Source |
|---|---|---|---|---|
| In-plane stiffness | $E_{2D}$ | 35 | N/m | [S13] (DFT) |
| Poisson ratio | ν | 0.42 | — | [S13] (DFT); directional range 0.40–0.44 |
| *h*-CuI lattice constant | $a_{CuI}$ | 4.19 | Å | [S10] (DFT) |
| Graphene superlattice | $\sqrt{3}\cdot a_G$ | 4.26 | Å | $a_G$ = 2.460 Å |
| Lattice mismatch | δ | 1.67 | % | $(\sqrt{3}\cdot a_G - a_{CuI})/a_{CuI}$ |
| Unit cell area | $A_{uc}$ | 15.2 | Å$^2$ | $\sqrt{3}/2 \cdot a_{CuI}^2$ (freestanding lattice; the cell contains two Cu and two I atoms) |
| Gr–CuI binding (direct DFT) | $E_{vdW}$ | 13.58 (pristine) / 13.65 (defective) | meV/Å$^2$ | our DFT (Methods); net cross-check: 14.1 [S10] |
| Gr–CuI corrugation | $\Delta U_P$ | 0.01–0.14 (central 0.07) | meV/Å$^2$ | DFT upper bound, Ullah et al. [S11] |

**Table S4.** Thermodynamic parameters used in the main-text energy balance.

| Parameter | Value | Source / method |
|---|---|---|
| γ → β phase penalty (bulk) ΔE | 3.0–3.4 (2×-bulk upper bound 6.6) | Yang et al. [S12] (bulk reference); conservative 2×-bulk upper bound for 2D correction |
| CuI/Gr corrugation $\Delta U_P$ | 0.01–0.14 (central 0.07) meV/Å$^2$ | DFT upper bound, Ullah et al. [S11] |
| Poisson ratio (*h*-CuI) ν | 0.42 ± 0.02 | Demirok et al. [S13] |

# S5. Analysis of the AIMD Trajectories

The AIMD trajectories analyzed here were produced with CP2K as described in the Methods section of the main paper, see Fig. S4. Each system was run as at least six restart segments of about 1.5 ps, at 300 and 600 K (Section S5b). Positions were cached every 5 fs, and the nuclear propagation time step was 0.5 fs. Velocities were obtained from the cached positions by central finite difference.

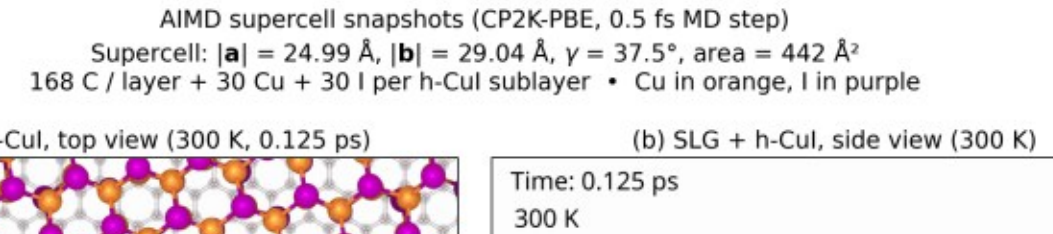

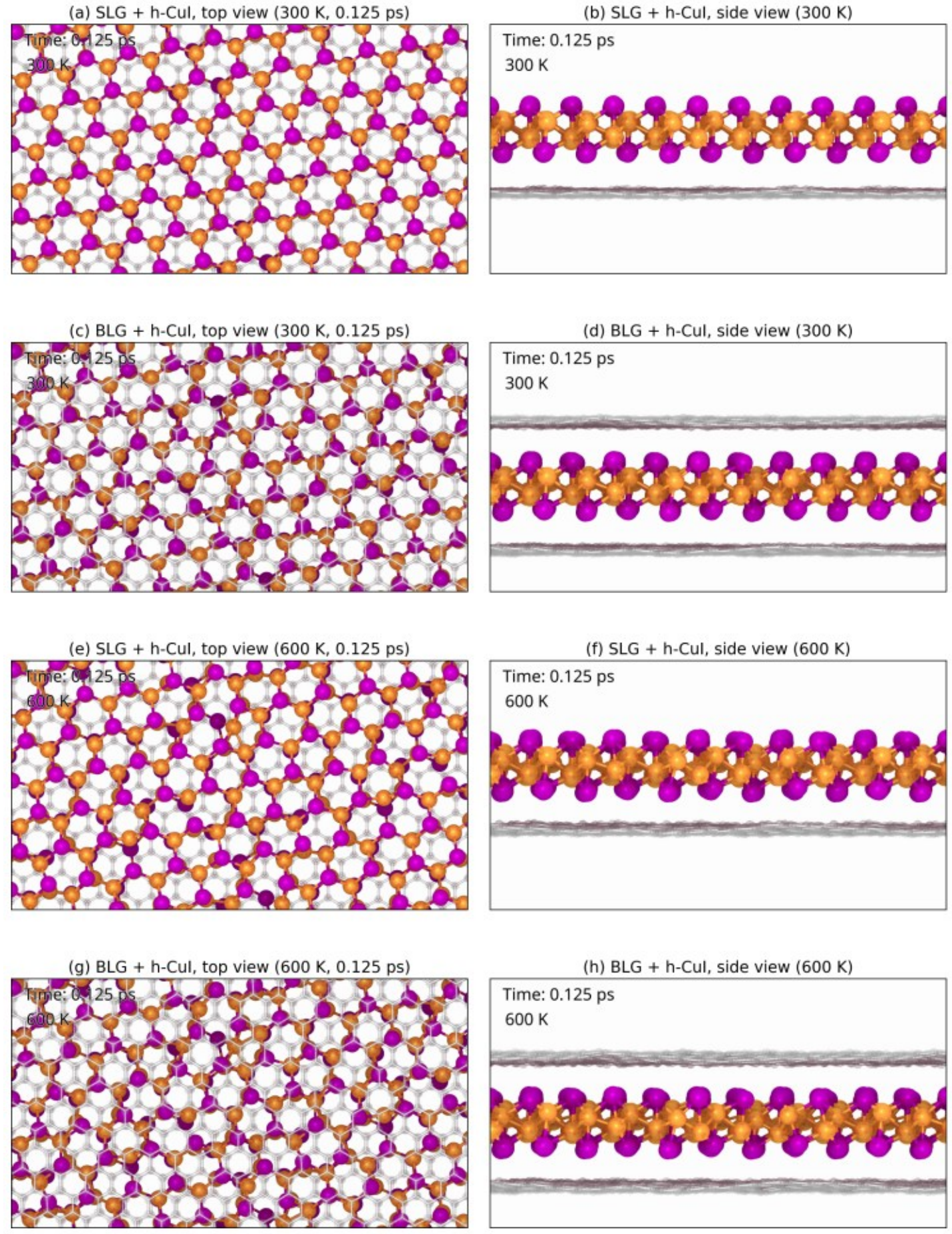

**Figure S4. AIMD supercell geometry.** Left column, top views; right column, side views. (a, b) *h*-CuI on single-layer graphene (SLG) at 300 K: the graphene lattice (gray) with overlaid *h*-CuI (Cu orange, I purple); the moiré stripe pattern is visible in the top view, and the side view shows one graphene layer on one side and vacuum on the other. (c, d) *h*-CuI within bilayer graphene (BLG) at 300 K: the mirror-symmetric I–Cu–Cu–I sandwich (bilayer thickness 3.93 Å) between two graphene layers. (e, f) SLG at 600 K and (g, h) BLG at 600 K, in the same views. Supercell vectors $|R_1|$ = 24.99 Å, $|R_2|$ = 29.04 Å, the angle between them γ = 37.5°, and the area $A_{cell}$ = 441.7 Å$^2$ are identical in all panels, as is the epitaxial twist angle $\theta = |30° - \Delta| \approx 20.6°$ between the *h*-CuI and the graphene lattice, with $\Delta \approx 9.4°$ the angular difference between the nearest lattice vectors (Section S2c); only the number of graphene layers and the temperature differ. (Supercell geometry at t = 0.125 ps (CP2K, PBE/GPW, nuclear propagation time step 0.5 fs).)

The cell holds six sublayers: the graphene layers $C_{bot}$ and $C_{top}$, the copper sublayers $Cu_{bot}$ and $Cu_{top}$, and the iodine sublayers $I_{bot}$ and $I_{top}$ of the I–Cu–Cu–I sandwich. The supported cell has no $C_{top}$. The collective coordinates read in Fig. 6 of the main paper are defined as follows.

- $Z_{anti} = \frac{1}{2}(\bar{z}_{top} - \bar{z}_{bot})$, the antisymmetric out-of-plane coordinate of the two CuI sublayers, formed from their center-of-mass heights.

- TO = $\bar{v}_y(Cu) - \bar{v}_y(I)$, the in-plane transverse optical coordinate, formed from the center-of-mass velocities of the copper and of the iodine sublattice as a whole.
- The coherence of Fig. 6(c) combines the center-of-mass height of a graphene layer with that of the adjacent CuI sublayer.

Power spectra were formed per restart segment. The coordinate was multiplied by a Hann window, zero-padded to 8192 samples and Fourier-transformed. The segment spectra were averaged with weights equal to their post-equilibration length, at least 200 frames each. The mean was then scaled to its own maximum, taken over 0–5.6 THz in Fig. 6(a) and over 1.5–6.5 THz in Fig. 6(b). Both spectra are plotted unsmoothed.

For the coherence the auto- and cross-spectra were accumulated with the same weights. The squared coherence $\gamma^2 = |S_{AB}|^2/(S_{AA}\, S_{BB})$ was formed after the average and smoothed with a 0.10 THz Gaussian. The band positions, spectral weights and coherences read from Fig. 6(a)–(c) are collected in Table S5.

In Fig. 6 of the main paper, a frequency is resolved in a segment-averaged spectrum only if every contributing segment holds at least one full period of it. As the individual segments give 1/T between 0.39 and 0.85 THz, the resolution limit of a run is set by the largest of these, 0.85 THz.

Mean-square displacements were averaged over time origins within each segment and then over segments. The plateau values of Fig. 6(i) are averages over lag times $\tau$ = 0.3–1.0 ps. The in-plane plateau values are collected in Table S6, and the top-over-bottom sublayer ratios for both species and both directions in Table S7. The out-of-plane displacements of Fig. 6(g) and 6(h) are resolved by sublayer after removing the center-of-mass motion of that sublayer, so they measure internal roughness. For a bound atom the plateau equals $2\sigma_z^2$, so the amplitude follows as $\sigma_z = \sqrt{(MSD/2)}$.

Fig. S5 shows the full frequency range, including the region below 0.9 THz, which carries less statistical weight. The supported *h*-CuI sheet follows the out-of-plane motion of its graphene layer more closely than either pair of the encapsulated cell.

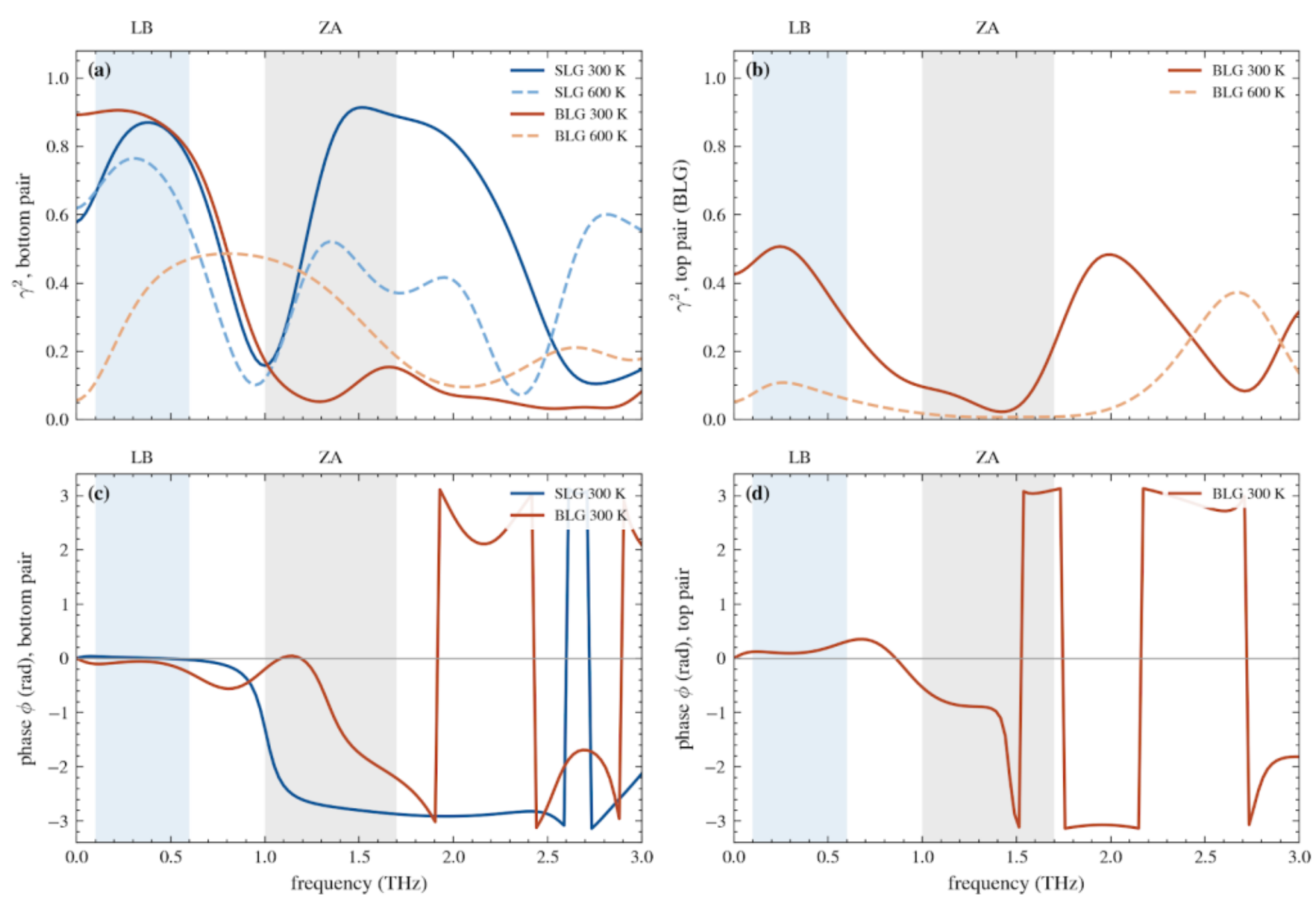

**Figure S5. Squared spectral coherence between the out-of-plane center-of-mass coordinate of a graphene layer and that of the adjacent CuI sublayer, for *h*-CuI on single-layer graphene and within bilayer graphene, with the bottom and the top pair shown separately**. Spectra are segment-averaged as described in Section S5a and smoothed with a 0.10 THz Gaussian; the region below 0.9 THz lies at or below the resolution limit of the individual restart segments and is shown for completeness only. (a) Bottom pair, bottom graphene against the bottom CuI sublayer, for all four systems. (b) Top pair, top graphene against the top CuI sublayer, encapsulated cell at both temperatures. Solid lines are 300 K and dashed lines 600 K. The shaded interval marks 1.0–1.7 THz, where the curvature of the adhesion potential of Eq. (2) of the main text places the interfacial breathing mode at 1.59 THz.

**Table S5. Spectral quantities read from Fig. 6(a)–(c), with uncertainties over the restart segments.** SLG is the supported cell and BLG the encapsulated cell. Spectral weight is the mean of the normalized power spectrum over the interval given. Band positions are maxima of the segment-averaged spectra, and the two positions of the internal Cu–I band agree within their uncertainties.

| **Quantity** | **SLG** | **BLG** |
|---|---|---|
| *$Z_{anti}$ main band (internal Cu–I), 300 K [THz]* | 3.93 ± 0.22 | 4.20 ± 0.08 |
| *In-plane band TO, maximum, 300 K [THz]* | 1.93 ± 0.21 | 2.76 ± 0.05 |
| *In-plane band TO, maximum, 600 K [THz]* | 1.86 ± 1.06 | 2.66 ± 1.23 |
| *$\gamma^2$, 0.9–3.0 THz, 300 K* | 0.525 ± 0.050 | 0.083 ± 0.134 (bottom pair), 0.206 ± 0.176 (top pair) |
| *$\gamma^2$, 1.0–1.7 THz, 300 K* | 0.656 ± 0.067 | 0.101 ± 0.317 (bottom pair), 0.070 ± 0.091 (top pair) |

**Table S6. In-plane mean-square displacement of copper and of iodine, averaged over lag times τ = 0.3–1.0 ps and over the restart segments (Fig. 6(f)).** SLG is the supported cell and BLG the encapsulated cell.

| **Species** | **SLG 300 K** | **SLG 600 K** | **BLG 300 K** | **BLG 600 K** |
|---|---|---|---|---|
| *Cu $MSD_{xy}$ [$Å^2$]* | 0.193 | 0.655 | 0.207 | 0.582 |
| *I $MSD_{xy}$ [$Å^2$]* | 0.085 | 0.220 | 0.093 | 0.169 |

**Table S7. Sublayer ratios of the mean-square displacement, top sublayer over bottom sublayer, for both species and both directions.** SLG is the supported cell and BLG the encapsulated cell. Each ratio is formed within one run. In the supported cell the bottom sublayer is in contact with graphene and the top faces vacuum. In the encapsulated cell both are in contact, so its columns are the control.

| **Quantity** | **SLG 300 K** | **SLG 600 K** | **BLG 300 K** | **BLG 600 K** |
|---|---|---|---|---|
| I, out of plane | 1.218 ± 0.048 | 1.301 ± 0.029 | 1.036 ± 0.030 | 0.952 ± 0.059 |
| I, in plane | 0.951 ± 0.063 | 0.934 ± 0.031 | 0.998 ± 0.046 | 0.978 ± 0.020 |
| Cu, out of plane | 0.903 ± 0.028 | 0.967 ± 0.050 | 1.046 ± 0.047 | 0.971 ± 0.070 |
| Cu, in plane | 1.170 ± 0.120 | 0.991 ± 0.036 | 0.955 ± 0.108 | 1.226 ± 0.188 |
| I $MSD_z$, bottom / top [$Å^2$] | 0.055 / 0.067 | 0.126 / 0.164 | 0.056 / 0.058 | 0.133 / 0.126 |
| Cu $MSD_z$, bottom / top [$Å^2$] | 0.081 / 0.073 | 0.248 / 0.240 | 0.076 / 0.079 | 0.242 / 0.235 |

The trajectories cover 300 and 600 K in one commensurate supercell, on restart segments of about 1.5 ps. Within these segments the iodine displacement reaches a plateau in both directions, while the copper in-plane displacement continues to rise. Amplitude comparisons between two separate runs carry the full segment scatter, and the copper in-plane difference between the supported and the encapsulated cell at 600 K stays within it. The quantities compared between the two cells in the main text are band positions, band-averaged spectral weights and window-averaged coherences, which are stable over the segments (Section S5b). Time scales beyond the restart segments are outside the reach of these trajectories.

## Supporting References